\documentclass[pra,aps,twocolumn,superscriptaddress,showpacs,nofootinbib,notitlepage]{revtex4-2}
\usepackage{graphicx}
\usepackage{amsmath,amssymb,amsthm,mathrsfs,amsfonts,dsfont}
\usepackage{subfigure, epsfig}
\usepackage{braket}
\usepackage{bm}
\usepackage[ruled,noline]{algorithm2e}
\usepackage{framed}
\usepackage{tcolorbox}
\usepackage{multirow}
\usepackage{verbatim}
\usepackage{diagbox}
\usepackage{enumerate}
\usepackage{physics} 
\usepackage{url}
\usepackage{color}
\usepackage{amscd}
\usepackage[normalem]{ulem}
\usepackage{pgfplots} 
\pgfplotsset{compat=1.17}
\usepackage{float}
\usepackage{tikzpeople}
\usepackage[10pt]{moresize}
\usepackage{hyperref}

\DeclareUnicodeCharacter{2212}{\textminus}
\newtheorem{theorem}{Theorem}
\newtheorem{lemma}{Lemma}

\newtheorem{definition}{Definition}

\newcommand{\Hmin}{H_{\text{min}}}

\newcommand{\Renyi}{\mathcal{R}_{\alpha}}

\usepackage{url}

\makeatletter
\let\saved@includegraphics\includegraphics
\AtBeginDocument{\let\includegraphics\saved@includegraphics}
\renewenvironment*{figure}{\@float{figure}}{\end@float}

\makeatother

\newcommand{\tabincell}[2]{\begin{tabular}{@{}#1@{}}#2\end{tabular}}

\begin{document}
\preprint{APS/123-QED}
\title{Device-Independent-Quantum-Randomness-Enhanced Zero-Knowledge Proof}

\author{Cheng-Long Li}
\affiliation{Hefei National Laboratory for Physical Sciences at Microscale and Department of Modern Physics, University of Science and Technology of China, Hefei 230026, P.~R.~China}
\affiliation{Shanghai Branch, CAS Center for Excellence and Synergetic Innovation Center in Quantum Information and Quantum Physics, University of Science and Technology of China, Shanghai 201315, P.~R.~China}
\affiliation{Shanghai Research Center for Quantum Sciences, Shanghai 201315, P. R. China}
	
\author{Kai-Yi Zhang}
\affiliation{Shanghai Jiao Tong University, Shanghai 200240, P.~R.~China}
	
\author{Xingjian Zhang}
\affiliation{Center for Quantum Information, Institute for Interdisciplinary Information Sciences, Tsinghua University, Beijing 100084, P.~R.~China}
	
\author{Kui-Xing Yang}
\affiliation{Shenzhen Institute for Quantum Science and Engineering and Department of Physics, Southern University of Science and Technology, Shenzhen, 518055, P.~R.~China}
	
\author{Yu Han} 
\affiliation{Center for Quantum Information, Institute for Interdisciplinary Information Sciences, Tsinghua University, Beijing 100084, P.~R.~China}
\affiliation{State Key Laboratory of Mathematical Engineering and Advanced Computing, Zhengzhou 450001, Henan, P.~R.~China}
	
\author{Su-Yi Cheng}
\affiliation{Hefei National Laboratory for Physical Sciences at Microscale and Department of Modern Physics, University of Science and Technology of China, Hefei 230026, P.~R.~China}
\affiliation{Shanghai Branch, CAS Center for Excellence and Synergetic Innovation Center in Quantum Information and Quantum Physics, University of Science and Technology of China, Shanghai 201315, P.~R.~China}
\affiliation{Shanghai Research Center for Quantum Sciences, Shanghai 201315, P. R. China}
	
\author{Hongrui Cui}
\affiliation{Shanghai Jiao Tong University, Shanghai 200240, P.~R.~China}

\author{Wen-Zhao Liu}
\affiliation{Hefei National Laboratory for Physical Sciences at Microscale and Department of Modern Physics, University of Science and Technology of China, Hefei 230026, P.~R.~China}
\affiliation{Shanghai Branch, CAS Center for Excellence and Synergetic Innovation Center in Quantum Information and Quantum Physics, University of Science and Technology of China, Shanghai 201315, P.~R.~China}
\affiliation{Shanghai Research Center for Quantum Sciences, Shanghai 201315, P. R. China}
	
\author{Ming-Han Li}
\affiliation{Hefei National Laboratory for Physical Sciences at Microscale and Department of Modern Physics, University of Science and Technology of China, Hefei 230026, P.~R.~China}
\affiliation{Shanghai Branch, CAS Center for Excellence and Synergetic Innovation Center in Quantum Information and Quantum Physics, University of Science and Technology of China, Shanghai 201315, P.~R.~China}
\affiliation{Shanghai Research Center for Quantum Sciences, Shanghai 201315, P. R. China}
	
\author{Yang Liu}
\affiliation{Jinan Institute of Quantum Technology, Jinan 250101, P.~R.~China}
	
\author{Bing Bai}
\affiliation{Hefei National Laboratory for Physical Sciences at Microscale and Department of Modern Physics, University of Science and Technology of China, Hefei 230026, P.~R.~China}
\affiliation{Shanghai Branch, CAS Center for Excellence and Synergetic Innovation Center in Quantum Information and Quantum Physics, University of Science and Technology of China, Shanghai 201315, P.~R.~China}
\affiliation{Shanghai Research Center for Quantum Sciences, Shanghai 201315, P. R. China}
	
\author{Hai-Hao Dong}
\affiliation{Hefei National Laboratory for Physical Sciences at Microscale and Department of Modern Physics, University of Science and Technology of China, Hefei 230026, P.~R.~China}
\affiliation{Shanghai Branch, CAS Center for Excellence and Synergetic Innovation Center in Quantum Information and Quantum Physics, University of Science and Technology of China, Shanghai 201315, P.~R.~China}
\affiliation{Shanghai Research Center for Quantum Sciences, Shanghai 201315, P. R. China}
	
\author{Jun Zhang}
\affiliation{Hefei National Laboratory for Physical Sciences at Microscale and Department of Modern Physics, University of Science and Technology of China, Hefei 230026, P.~R.~China}
\affiliation{Shanghai Branch, CAS Center for Excellence and Synergetic Innovation Center in Quantum Information and Quantum Physics, University of Science and Technology of China, Shanghai 201315, P.~R.~China}
\affiliation{Shanghai Research Center for Quantum Sciences, Shanghai 201315, P. R. China}
	
\author{Xiongfeng Ma}
\affiliation{Center for Quantum Information, Institute for Interdisciplinary Information Sciences, Tsinghua University, Beijing 100084, P.~R.~China}
	
\author{Yu Yu}
\affiliation{Shanghai Jiao Tong University, Shanghai 200240, P.~R.~China}
	
\author{Jingyun Fan}
\affiliation{Shenzhen Institute for Quantum Science and Engineering and Department of Physics, Southern University of Science and Technology, Shenzhen, 518055, P.~R.~China}
	
\author{Qiang Zhang} 
\affiliation{Hefei National Laboratory for Physical Sciences at Microscale and Department of Modern Physics, University of Science and Technology of China, Hefei 230026, P.~R.~China}
\affiliation{Shanghai Branch, CAS Center for Excellence and Synergetic Innovation Center in Quantum Information and Quantum Physics, University of Science and Technology of China, Shanghai 201315, P.~R.~China}
\affiliation{Shanghai Research Center for Quantum Sciences, Shanghai 201315, P. R. China}
\affiliation{Jinan Institute of Quantum Technology, Jinan 250101, P.~R.~China}
	
\author{Jian-Wei Pan}
\affiliation{Hefei National Laboratory for Physical Sciences at Microscale and Department of Modern Physics, University of Science and Technology of China, Hefei 230026, P.~R.~China}
\affiliation{Shanghai Branch, CAS Center for Excellence and Synergetic Innovation Center in Quantum Information and Quantum Physics, University of Science and Technology of China, Shanghai 201315, P.~R.~China}
\affiliation{Shanghai Research Center for Quantum Sciences, Shanghai 201315, P. R. China}

\begin{abstract}

Zero-knowledge proof (ZKP) is a fundamental cryptographic primitive that allows a prover to convince a verifier of the validity of a statement without leaking any further information\cite{goldwasser1989knowledge}. As an efficient variant of ZKP, non-interactive zero-knowledge proof (NIZKP) adopting the Fiat-Shamir heuristic is essential to a wide spectrum of applications, such as federated learning\cite{yang2019federated}, blockchain and social networks\cite{kappos2018empirical}. However, the heuristic is typically built upon the random oracle model making ideal assumptions about hash functions, which does not hold in reality and thus undermines the security of the protocol\cite{canetti2004random}. Here, we present a quantum resolution to the problem. Instead of resorting to a random oracle model, we implement a quantum randomness service. This service generates random numbers certified by the loophole-free Bell test\cite{bierhorst2018experimentally,liu2018device} and delivers them with post-quantum cryptography (PQC) authentication\cite{zhang2020tweaking}. Employing this service, we conceive and implement a NIZKP of the three-colouring problem\cite{goldreich1991proofs}.
By bridging together three prominent research themes, quantum non-locality,
PQC and ZKP, we anticipate this work to open a new paradigm of quantum information science. 
\end{abstract}

\maketitle

From bank loans to adding friends on social networks, while sharing our personal data to validate ourselves, we might raise the following doubt: is it really necessary to share so much information?
To unveil nothing more than exactly required by the tasks, we can use ZKP\cite{kappos2018empirical}. In the original format of ZKP, a prover needs to interact with a verifier for many rounds of challenges. 
In multi-party applications, unfortunately, such a one-to-one highly interactive protocol becomes impractical or even unrealisable.
In comparison, NIZKP sends a one-round message to convince multiple verifiers, hence is favourable in real-life usage.
NIZKP typically relies on the Fiat-Shamir heuristic that is instantiated in the random oracle model. The outcome of cryptographic hash functions, such as SHA2 and SHA3\cite{pub2012secure,dworkin2015sha}, are used as cryptographic random numbers in the random oracle model with a known input. Though the hash functions based on high computational complexity make the pseudo-random numbers practically hard to guess, the hashing functions are deterministic in essence. On the one hand, no intrinsic randomness can be produced from a deterministic function. A few insecure counterexamples have been found about the use of the random oracle model\cite{canetti2004random}.
On the other hand, the belief in computational complexity is challenged by the emerging quantum computing technology which promises unprecedented computing power\cite{arute2019quantum,zhong2020quantum}.

Inspired by the principle of quantum non-locality\cite{bell1964einstein}, we show that quantum physics offers a resolution to the problem.
We report in this work a quantum randomness service that consists of a randomness beacon and a timestamp server.
The randomness beacon produces random numbers from loophole-free Bell tests, namely device-independently, and broadcasts them to the public with PQC-based algorithms.
As the entropy source, the device-independent quantum random number generation (DIQRNG) does not require any prior assumptions or characterisations of the inner working status of the quantum devices. Instead, the generated randomness is certified based solely on the quantum nonlocal behaviour\cite{colbeck2009quantum}. For the delivery of random numbers, we apply a lattice-based signature algorithm that is secure against known quantum algorithms. In addition, we also provide a timestamp server, which signs the digital message on requirement with PQC-based algorithms. The combination of DIQRNG and PQC signature bring a high-security level for the public randomness service.
Based on this service, we conceive and experimentally demonstrate an efficient and secure NIZKP of NP-complete three-colouring problem. In the following, we first present a NIZKP protocol of NP-complete three-colouring problem, then introduce its experimental realisation, central to which is the preparation of a quantum randomness service.

\textbf{Results. ---}Graph colouring is a famous problem in computational complexity.
Formally, a graph $G(V,E)$ with vertices $V$ connected by edges $E$ is three-colourable if there exists a mapping $\phi : V \rightarrow \{1,2,3\}$ such that every two adjacent vertices connected by an edge have different colours, i.e., $\forall u, v \in V$ and $(u,v)\in E, \phi(u)\neq \phi(v)$.
As a special case of graph colouring, three-colouring is $\mathsf{NP}$-Complete\cite{arora2009computational}.
That is, each $\mathsf{NP}$ problem instance corresponds to an instantiation of a particular three-colourable graph. Hence a secure and efficient protocol to prove the three-colourability is of fundamental interest. In our NIZKP of three-colouring problem, the prover can convince that he has a solution to the problem without revealing the specific colouring assignments.

\begin{figure*}
\centering
\includegraphics[width = 0.7\textwidth]{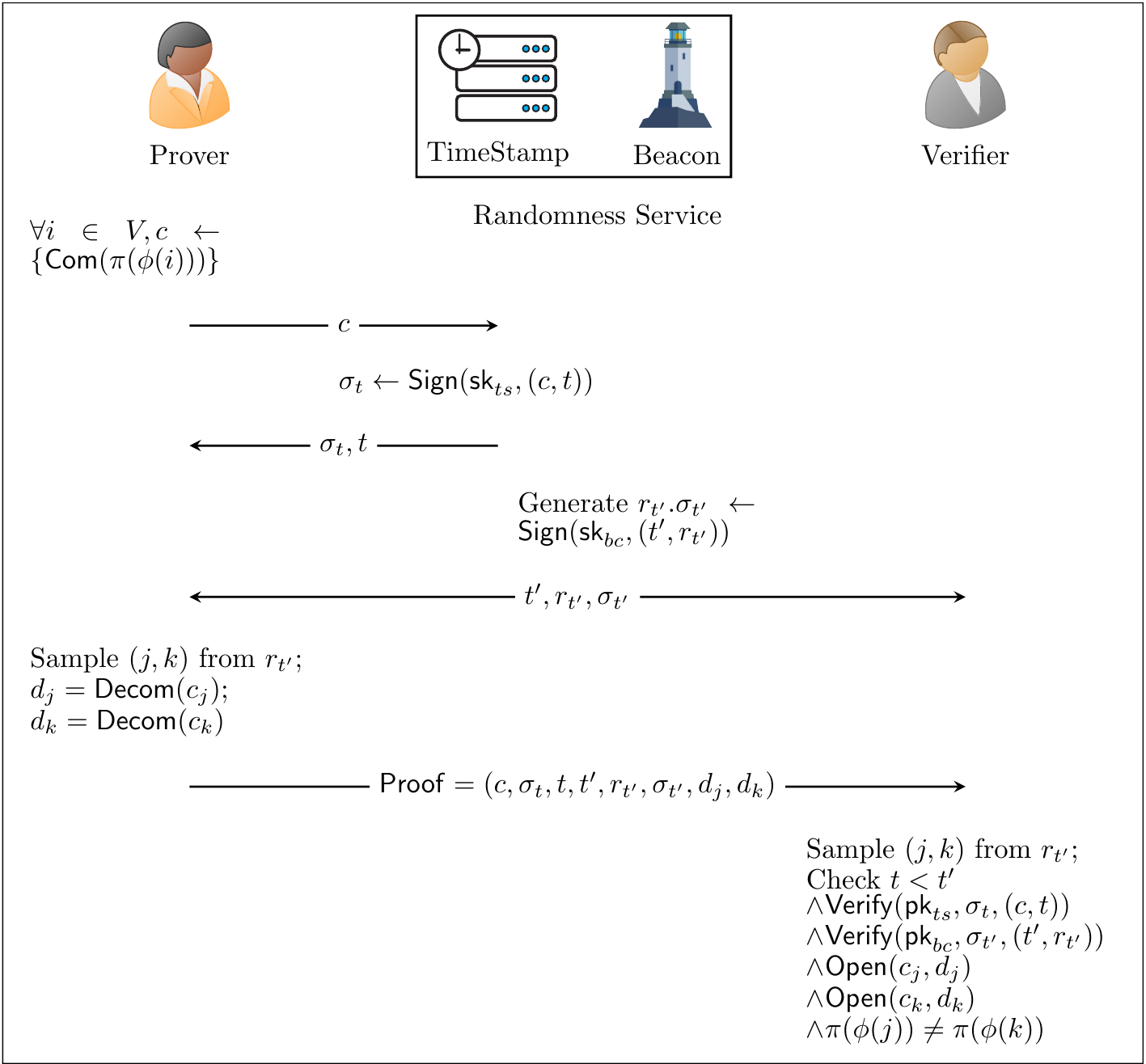}
\caption{A NIZKP of three-colourability based on a randomness service. With the aid of the randomness service that is equipped with a beacon and timestamp, the prover hopes to prove the knowledge of three-colourability to the verifier in the fashion of NIZKP. We depict the computation and communication steps in time order. Notations are defined by the following. A commitment scheme consists of two algorithms, ${\sf Com}$ and ${\sf Decom}$. ${\sf Com}$ on the inputs of a message outputs a commitment and instantiated by the SHA256 hash function in this experiment. ${\sf Decom}$ on the inputs of a commitment outputs a decommitment. 
Open on the inputs of a commitment and a decommitment, outputs false or true, corresponding whether the commitment and the decommitment are consistent. Here, we verify if the decommitment is the key of the SHA256 hash function.  {$\sf Sign$} on the inputs of a secret key and a message outputs a signature. {$\sf Verify$} on the inputs of public keys, a signature and a message output false or true, corresponding to whether the message is signed by the owner of the public key. {$\sf Proof$} contains all the information that is required for the verification. }.

\label{fig:nizkp generation 2}
\end{figure*}

Our protocol of the three-colouring problem is compiled from the basic sigma protocol of the classic three-colouring protocol\cite{goldwasser1989knowledge}. Different from the common NIZKP protocol, apart from the prover and verifier, we introduce a randomness service, as shown in Fig.~\ref{fig:nizkp generation 2}. With the aid of the service, the prover and verifier remain the capability to generate a non-interactive proof directly, while they no longer need to resort to the random oracle assumption. The randomness service is essentially composed of a randomness beacon\cite{rabin1983transaction} and a timestamp server. The random service must be secure and the prover needs to show that the challenge from the beacon is generated \emph{after} the commitment is published. With a trusted timestamp server, the commitment can be certified at the time of the event.

Initially, the timestamp server and the beacon each generate a pair of public key and secret key with PQC denoted by  $(pk_{ts},sk_{ts})$ and $(pk_{bc},sk_{bc})$, respectively, keep the secret key and share the public key with all participants. 
We then execute the protocol as follows:
\begin{enumerate}
\item The prover prepares an assignment of colouring denoted by $\phi$, chooses uniformly a permutation $\pi \in S_3$, commits $\pi(\phi(i))$ for all vertices  $i\in V$ which are denoted by $c$, $\{\text{Com}(\pi(\phi(i)))\}\rightarrow c$, and sends $c$ to the timestamp server.
\item The timestamp server signs $c$ and the received time $t$ on a signature $\sigma_t$, $\text{Sign}(sk_{ts},c,t)\rightarrow \sigma_t$, sends $\sigma_t$ and $t$ to the prover.
\item  A beacon broadcasts the random bits $r_{t'}$ with signature $\sigma_{t'}$, $\text{Sign}(sk_{bc}, t',r_{t'})\rightarrow \sigma_{t'}$.
\item The prover chooses a random edge $e=(j,k)\in E$ according to $r_{t'}$, and decommits $c_j,c_k$ to obtain $d_j=\text{Decom}(c_j)$, $d_k=\text{Decom}(c_k)$.
\item The prover sends the final proof $(c, \sigma_t, t, t', r_{t'},\sigma_{t'},e,d_j,d_k)$ to the verifier.
\item The verifer checks if the commitments, the random bits, and the signature are correct, and the colour of the two vertices are different.
\end{enumerate}

Completeness of this protocol follows inspection. For protocol soundness, if the graph is not three-colourable, then at least one edge has the same colour on both ends (vertices). Assuming the commitment protocol is perfectly binding, the protocol soundness error is at most $(|E|-1)/|E|$  and can be boosted to $2^{-\lambda}$ for an arbitrary parameter $\lambda>0$ by repeating the process\cite{arora2009computational}. The zero-knowledge property follows from the hiding property of the commitment protocol.

\begin{figure*}[htbp]
\centering
\subfigure[]
{\centering
\includegraphics[width = 0.7\textwidth]{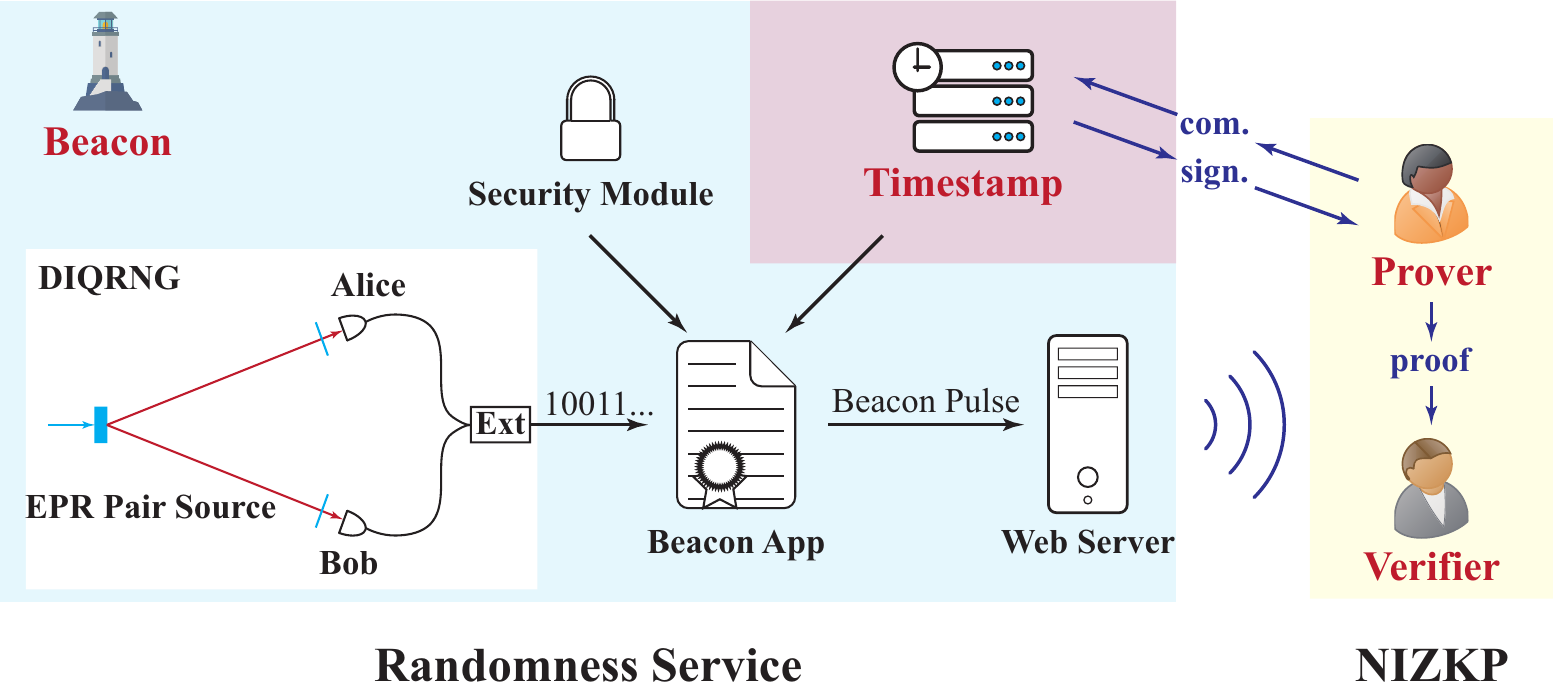}
\label{fig:Beacon_Structure}}
\subfigure[]
{
\includegraphics[width = 0.53\textwidth]{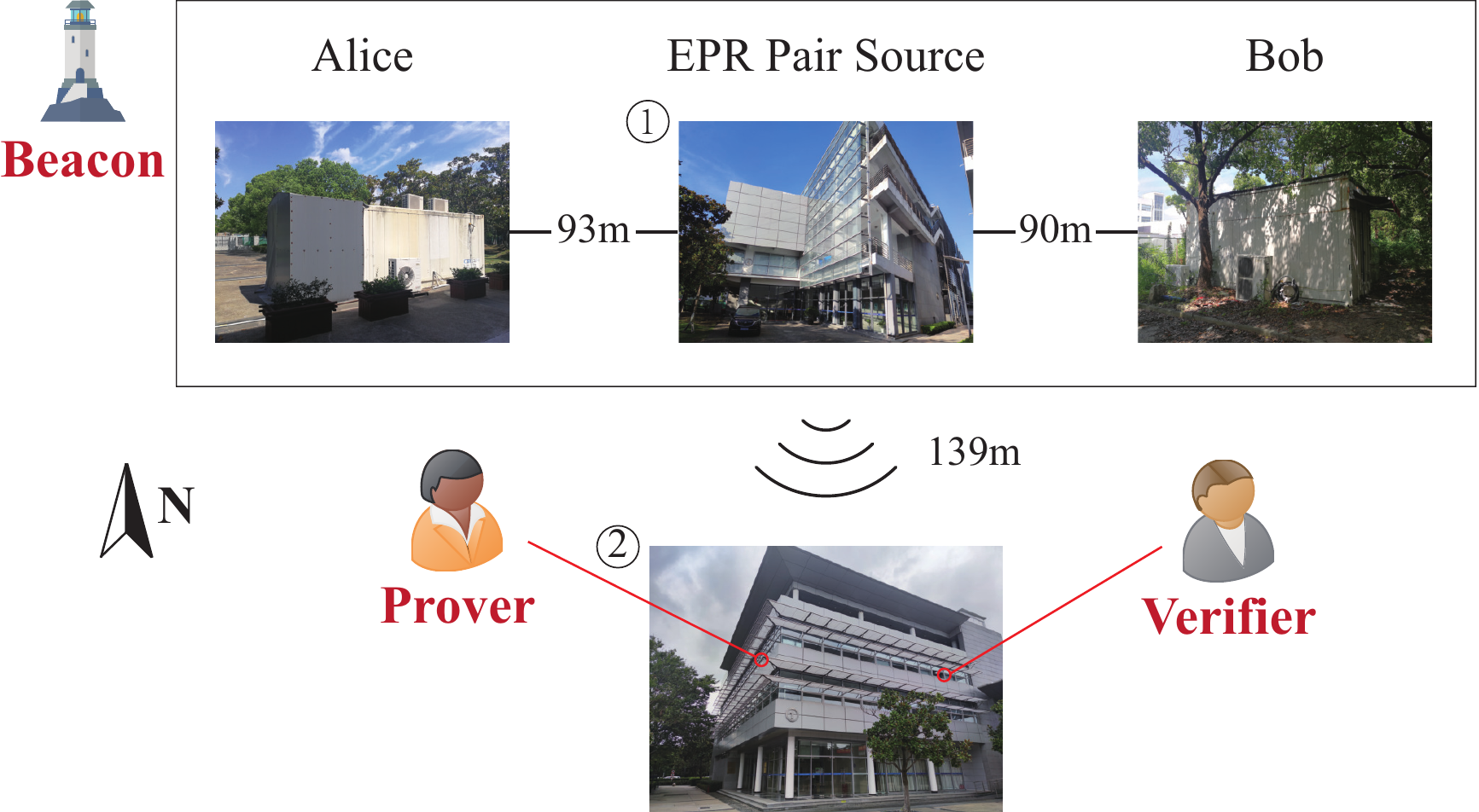}
\label{fig:real_scene}}
\subfigure[]
{
\includegraphics[width = 0.4\textwidth]{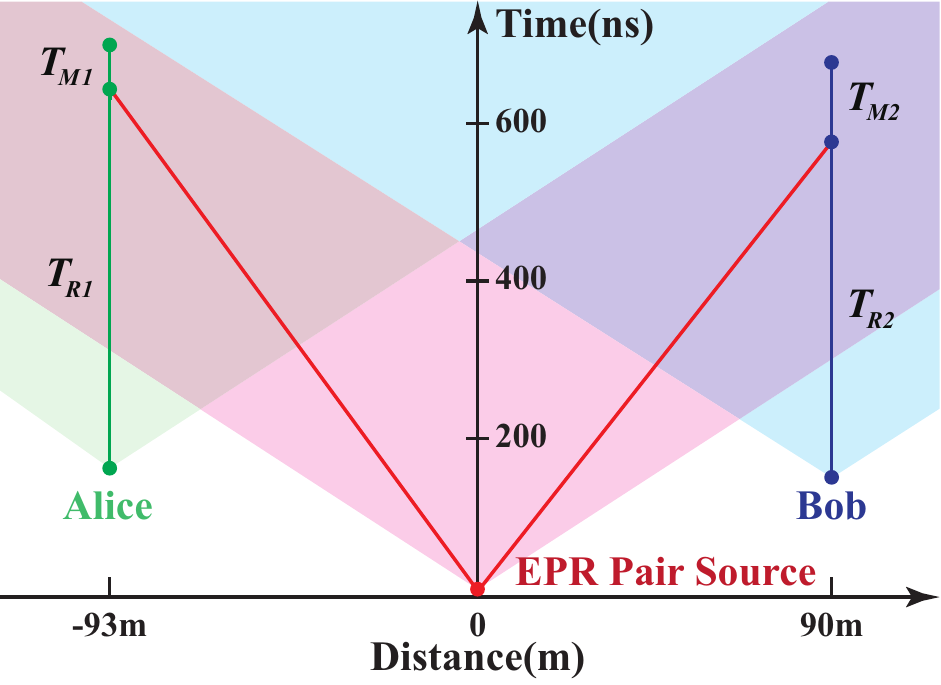}
\label{fig:DI_Setup}}
\caption{Schematics of the experiment. (a) A flowchart demonstration of the experiment. The randomness service contains Randomness Beacon and Timestamp Server. The randomness beacon broadcasts random bits and the timestamp server signs the commitment and the received time. Prover and Verifier communicate with the beacon and timestamp to carry out the NIZKP protocol. The timestamp server is synchronised with Coordinated Universal Time (UTC). The beacon contains a device-independent random number generator (DIQRNG) as the entropy source, a Security Module to provide the authenticity of signed beacon pulses, a Beacon APP to process inputs from DIQRNG, a timestamp from Timestamp Server, and the Web Server broadcasts the beacon pulse containing 512 bits to the public.
(b) Schematic diagram of the overall experimental set-up. An EPR pair source is located in Building 1, where pairs of entangled photons are transmitted through a fibre to the measurement stations in opposite directions with free-space distances of 93 m and 90 m, denoted as Alice and Bob, respectively. Other beacon components are located in the same lab as the EPR source. The prover and the verifier are located in Building 2, which is 139 m from Building 1. (c) Space-time diagram for the DIQRNG experimental configuration. The values $T_{R1,2}$ are the time for Pockels cells to get ready for state measurement and $T_{M1,2}$ are the time elapse for single-photon detectors to output an electronic signal.
No signalling between relevant events is allowed in this configuration (See Appendix \ref{sec:spacetime_details} for more details)}.	
\label{fig:Beacon_Structure_DIQRNG_NIZKP}
\end{figure*}

We implement the randomness-service-enhanced NIZKP protocol, with the flow-chart shown in Fig.~\ref{fig:Beacon_Structure_DIQRNG_NIZKP}(a).
In this proof-of-principle experiment, the randomness beacon broadcasts timed outputs of fresh public randomness in a block every minute continuously. A block, which is called a `beacon pulse', contains 512 fresh random bits that are time-stamped, signed, and hash-chained\cite{fischer2011public}.
The timestamp server keeps waiting for the requirement of signing the commitment. The prover sends the proof to the verifier. Then, the verifier opens the signature and commitment and checks if the conditions are satisfied.

For the security of the protocol, the randomness beacon needs to satisfy a set of properties required for a high level of trust: (1) Unpredictability: the values of random bits cannot be predicted before their generation; (2) Autonomy: the source is resistant to any outside attempt to alter the distribution of random bits; (3) Consistency: a set of users can receive the same random bit-strings from the broadcast\cite{fischer2011public}.

In the construction of the randomness beacon, we adopt the DIQRNG as the entropy source\cite{pironio2010,bierhorst2018experimentally,liu2018device,shalm2021device,li2021experimental}. The device-independent feature of DIQRNG guarantees the requirement of unpredictability. In essence, we realise a loophole-free Clauser-Horne-Shimony-Holt-type (CHSH) Bell test experiment\cite{Hensen_Loophole_2015,Shalm15,Giustina15,Rosenfeld17,LiPRL2018}.
As shown in Fig.~\ref{fig:real_scene}, an Einstein-Podolsky-Rosen (EPR) state source in Building 1 emits a pair of photons in the EPR state and delivers one photon to Alice at a distance of 93 m to the west and another photon to Bob at a distance of 90 m to the east.
The spatial separations are sufficiently large to close the locality loophole. In Fig.~\ref{fig:DI_Setup}, we show the time-space coordinates of the relevant events.
The total system efficiency for photon delivery and detection reaches $81\%$, which is sufficient to close the detection loophole\cite{liu2018device,li2021experimental}.
We use the quantum probability estimation method to evaluate the randomness generated in the Bell test experiment\cite{zhang2020efficient} and utilise a quantum-proof strong randomness extractor\cite{ma2013postprocessing} to extract uniformly distributed random bits. 
The Bell test experiment generates 512 bits of randomness in about 31.38 seconds with a soundness error of $2^{-64}$. The randomness extraction takes approximately one second, with a soundness error of $2^{-100}$.

To fulfil autonomy and consistency, in the authentication procedure, we apply a lattice-based AIGIS PQC signature algorithm\cite{zhang2020tweaking} in the security module of the beacon to identify the randomness server. In addition, we apply in parallel an RSA-4096 signature algorithm\cite{rivest1978method}. The lattice-based signature algorithm is secure against all known quantum algorithms, providing quantum-safe protection. The random numbers needed to produce the public and private keys for RSA 4096 and Aigis PQC comes from our previous DIQRNG experiment\cite{li2021experimental}. The private key ($sk_{bc}$) is used to sign the message and the public key ($pk_{bc}$) is broadcast to everyone. The double certification ensures the trust of the randomness beacon service.

In addition to the randomness beacon, in our randomness service, we also provide a timestamp server synchronized to the Coordinated Universal Time (UTC). The timestamp server applies a lattice-based AIGIS PQC signature algorithm to provide the public key and private key ($pb_{ts}$, $sk_{ts}$) on the requirement. In implementing our NIZKP protocol, this timestamp server authenticates the commitment of the prover in each round of the experiment.

In implementing the NIZKP protocol, we arrange the prover and the verifier in Building 2. They receive the random bits broadcast by the beacon and are 139 m away from Building 1, as shown in Fig.~\ref{fig:real_scene}.
With the environment of both the prover and the verifier in OS Win10, CPU i7-9750H CPU @ 2.60GHz, RAM 16.0 GB, we extend the randomness $r$ by pseudorandom number generator on sampling the edge $e$ with SHA256 in counter mode. We generate randomness used in commitment and signature by the Intel chip command `{$\sf rdseed$}'. 

We proceed with the experimental demonstrations of NIZKP of the three-colouring problem with graphs of different sizes. With the number of edges $E$ set to $E=3V$, our experimental results show that the consumption of the resource, i.e., time of commit phase, time of response phase, time of verifying phase, and proof size, exhibits an asymptotic complexity of $\mathcal{O} (|V|^2)$ with respect to the number of vertices $V$, which complies with our algorithm, as shown in Table I. The overall soundness error of our protocol is the sum of $2^{-64}$ (randomness generation), $2^{-100}$ (randomness extraction), and $2^{-64}$ (ZKP).

\begin{table*}[t]
	\centering	
	
	\begin{tabular}{|l|l|l|l|l|l|l|}
		\hline
		$V$  & commit time/s & response time/s & verify time/s & proof size/MB & \# of rounds \\ \hline
		50  & 2.19       & 1.25         & 0.08          & 21.73         & 6632    \\ \hline
		100 & 8.72       & 4.92         & 0.23          & 84.03         & 13286   \\ \hline
		150 & 19.78       & 11.16         & 0.45          & 188.97        & 19940   \\ \hline
		200 & 35.52      & 19.65         & 0.73          & 334.46        & 26595   \\ \hline
		250 & 54.93      & 31.03         & 1.08          & 521.20        & 33249   \\ \hline
	\end{tabular}
	\caption{Results of our non-interactive zero-knowledge proof. Commitment time is the time consumption of the prover in step 1. Response time is the time consumption of the prover in step 2, 3 , 4 and 5. Verify time is the time consumption of the verifier in step 6. Note that we count only local computation time, except communication time. Proof size is the disk and communication consumption of the proof in step 5. The number of rounds is the number of parallel repetitions. The consumption of random service is fairly small, thus we omit them here.}
	\label{table:table1}
\end{table*}

\textbf{Conclusions. ---}We propose a NIZKP protocol with the aid of a public randomness service, which  removes the random oracle model. As a demonstration, we implement the protocol for the three-colouring problem. To realise the protocol, we build the randomness beacon, the core of our randomness service, using a loophole-free DIQRNG and quantum-safe signature algorithms, which enhance the security of the ZKP protocol. To our best knowledge, this is the first time that DIQRNG is applied to the randomness beacon. The frame of our randomness beacon forms a strong trust chain. The basic trust originates from the correctness of quantum mechanics, which guarantees the intrinsic unpredictability of random numbers. For the delivery of the random numbers, thanks to the use of quantum-safe authentication algorithms, the users can trust their access. Valid users can also verify the broadcast random numbers using the timestamps and signatures. In the future, the trust chain can be further boosted. By combining various randomness beacon services owned by separate administrative identities and even more countries in a quantum network, the users can bypass the problem of one or a few malicious beacon services. Starting from the fundamental discussion on local realism and quantum non-locality by Einstein, Podolsky, and Rosen in 1935\cite{einstein1935can}, continuous experimental and theoretical developments have innovated the most secure information-technology applications thus far, the device-independent quantum cryptography, and shall inspire more device-independent quantum information processing applications.

\textbf{Acknowledge. ---}This work was supported by the National Key Research and Development (R\&D)Plan of China (2018YFB0504300,2020YFA0309701,2017YFA0303900), the National Natural Science Foundation of China(T2125010),  the Chinese Academy of Sciences, the Anhui Initiative in Quantum Information Technologies, Shanghai Municipal Science and Technology Major Project (2019SHZDZX01), and the leading talents of Quancheng industry.

C.-L. L and K.-Y. Z. contributed equally to this work.


\appendix
\onecolumngrid
\section{Zero-knowledge proof for the three-colouring problem}\label{sec:zkp}

The  well-known three-colouring problem is NP-complete, which means that any NP problem can be converted to it in polynomial time~\cite{arora2009computational}. Therefore, a ZKP protocol can apply to all NP problems if it is applicable to the three-colouring problem. We first give the definition of three-colourability.

\begin{definition}
 A graph $G$ is three-colourable if the vertices of a given graph can be coloured with only three
colours, such that no two vertices of the same colour are connected by an edge.
\end{definition}

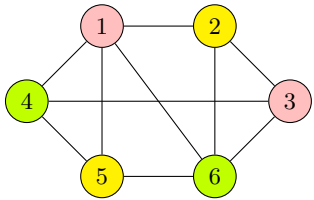
\begin{figure}[!ht]
	\centering
	
	\begin{tikzpicture}
		\node[shape=circle,draw=black,fill=pink] (1) at (0,0) {1};
		\node[shape=circle,draw=black,fill=yellow] (2) at (1.5,0) {2};
		\node[shape=circle,draw=black,fill=pink] (3) at (2.5,-1) {3};
		\node[shape=circle,draw=black,fill=lime] (4) at (-1,-1) {4};
		\node[shape=circle,draw=black,fill=yellow] (5) at (0,-2) {5};
		\node[shape=circle,draw=black,fill=lime] (6) at (1.5,-2) {6} ;
		
		\path [-] (1) edge (2); 
		\path [-] (1) edge (4); 
		\path [-] (1) edge (5); 
		\path [-] (1) edge (6); 
		\path [-] (2) edge (3); 
		\path [-] (2) edge (6); 
		\path [-] (3) edge (4); 
		\path [-] (3) edge (6); 
		\path [-] (4) edge (5); 
		\path [-] (5) edge (6); 
	\end{tikzpicture}
	
	\caption{An illustration of a three-colourable graph. In the showcase, with a colouring of only three colours, namely, pink, yellow, and green, every two adjacent vertices linked by an edge have different colours.}
	\label{fig:threecolor}

\end{figure}

For the three-colouring problem, there exists a zero-knowledge-proof protocol, where a prover can prove the statement about a specific graph's three-colourability to a verifier without revealing any knowledge~\cite{goldwasser1989knowledge}. 

Let $G$ be a graph of n vertices and define $V = \{1, ..., n\}$ be the set of vertices,
$E$ is the set of edges, $\phi(i)$ be the colour of $i$. On input the graph $G$ is known to both parties. The prover is given a private input in the protocol that is the witness which is a three-colouring of the graph $G$. The protocol proceeds as follows.
\begin{tcolorbox}[title = {Zero-knowledge proof for the three-colouring problem}]
\begin{enumerate}
    \item Prover: Randomly permute the three colours to obtain a new colouring. Utilize a commitment scheme to commit the colour of all vertices. $\forall i \in V, c \leftarrow \{{\sf Com}(\pi(\phi(i)))\}$. And then send $c$ to Verifier.
    \item Verifier: Generate random number $r$ and send it to Prover.
    \item Prover: Sample $ (j, k)$ from $ r$; $d_j = {\sf Decom}(c_j)$; $d_k = {\sf Decom}(c_k)$. Send $d_j,d_k$ to Verifier.
    \item Verifier: Accept if $\pi(\phi(j))\neq \pi(\phi(k))$ and open commitment correctly. Reject otherwise.
\end{enumerate}
\end{tcolorbox}

In the format of ZKP, it usually needs several communication steps to complete the protocol. However, Fiat-Shamir heuristic can transform the protocol from interactive to non-interactive by assuming the existence of the random oracle~\cite{fiat1986prove}, so that the protocol can be finished in one communication step.

\begin{definition}[Random Oracle\cite{10.1145/1008731.1008734}]
 In cryptography, a random oracle is an oracle (a theoretical black box) that responds to every unique query with a (truly) random response chosen uniformly from its output domain. If a query is repeated, it responds the same way every time that query is submitted.
\end{definition}

When we want to implement the scheme in the real world, a random oracle is not available. Instead, random oracle is widely used to model hash functions, although the outputs of hash functions are actually determined (without any unpredictability). Therefore the random oracle is oversimplified.  Most of Fiat Shamir Heuristic can be proved secure only in the random oracle model.

Let $m_1, m_2, m_3$ be the three messages sent between the Prover and the Verifier. The Fiat-Shamir transformation is executed as follows:
\begin{tcolorbox}[title = {Fiat-Shamir Transformation}]
 \begin{enumerate}
     \item The Prover sends the message $m_1$ to the Verifier. 
     \item The message $m_2$ is sent to the Prover based on the random picking of an edge which is viewed as a string.  $m_2$ is computed as $m_2=H(m_1)$. Where $H(\cdot)$ is a random oracle.
     \item Since $H(\cdot)$ is a public hash function, the computation can actually be done by the Prover. Thus the Prover sends $m_1, H(m_1), m_3$ to the Verifier.
 \end{enumerate}
\end{tcolorbox}

\begin{figure}
	\subfigure[]{
		\fbox{
			\begin{tikzpicture}[
				>=stealth,
				player/.style={
					minimum width=1cm
				},
				txt/.style={
					text width=3cm,
					anchor=north,
					execute at begin node=\setlength{\baselineskip}{3pt},
					scale=1,
					yshift=.3cm
				},
				msg/.style={
					fill=white,
					midway,
				},
				arrow/.style={
					->,thick,
					anchor=north}
				]
				\node (p) [player,alice] at (0,0) {};  
				\node (v) [player,bob,anchor=west] at ($(p.east)+(3.5,0)$) {};
				
				\node (plabel) [] at ($(p)-(0,1)$) {Prover};  
				\node (vlabel) [] at ($(v |- plabel)$) {Verifier};
				
				\node (pcom) [txt] at ($(p)-(0,2)$) {$\forall i\in V,c \gets \{{\sf Com}(\pi(\phi(i)))\}$};
				
				\draw  [->,thick] ($(p |- pcom.south)-(0,.5)$) -- ($(v |- pcom.south)-(0,.5)$) node (cmsg) [midway,msg] {$c$};
				
				\node (vrand) [txt] at ($(v |- cmsg.south)-(0,0.5)$) {Generate $r$};
				
				\draw [->,thick] ($(v |- vrand.south)-(0,.5)$) -- ($(p |- vrand.south)-(0,.5)$) node (vrandmsg) [midway,msg] {$r$};
				
				\node (pprove) [txt] at ($(p |- vrandmsg.south)-(0,.5)$) {
					Sample $ (j, k)$ from $ r$;\\ $d_j = {\sf Decom}(c_j)$;\\ $d_k = {\sf Decom}(c_k)$};

				\draw [->,thick] ($(p |- pprove.south)-(0,.5)$) -- ($(v |- pprove.south)-(0,.5)$) node (proof) [msg] {${\sf Proof} = (c, d_j,d_k)$};

				\node (vverify) [txt] at ($(v |- proof.south)-(0,0.75)$) {${\sf Open}(c_j,d_j)$  \\
					$\land {\sf Open}(c_k,d_k)$  \\
					$\land \pi(\phi(j))\neq \pi(\phi(k))$};
	\end{tikzpicture}}}
	\subfigure[]{\fbox{
			\begin{tikzpicture}[
				>=stealth,
				player/.style={
					minimum width=1cm
				},
				txt/.style={
					text width=3cm,
					anchor=north,
					execute at begin node=\setlength{\baselineskip}{3pt},
					scale=1,
					yshift=.3cm
				},
				msg/.style={
					fill=white,
					midway,
				},
				arrow/.style={
					->,thick,
					anchor=north}
				]
				\node (p) [player,alice] at (0,0) {}; 
				
				\node (v) [player,bob,anchor=west] at ($(p.east)+(3.5,0)$) {};
				
				\node (plabel) [] at ($(p)-(0,1)$) {Prover}; 
				\node (vlabel) [] at ($(v |- plabel)$) {Verifier};
				
				\node (pcom) [txt] at ($(p)-(0,2)$) {$\forall i\in V,c \gets \{{\sf Com}(\pi(\phi(i)))\}$};
				\node (rand) [txt] at ($(p)-(0,4)$) {{\color{red} $r=H(c)$;}};
				\node (pprove) [txt] at ($(p)-(0,5)$) {
					Sample $ (j, k)$ from $ r$;\\ $d_j = {\sf Decom}(c_j)$;\\ $d_k = {\sf Decom}(c_k)$};

				\draw [->,thick] ($(p |- pprove.south)-(0,0.5)$) -- ($(v |- pprove.south)-(0,0.5)$) node (proof) [msg] {${\sf Proof} = (c, d_j,d_k)$};

				\node (vverify) [txt] at ($(v |- proof.south)-(0,0.5)$) {$r=H(c)$\\ 
					$\land {\sf Open}(c_j,d_j)$  \\
					$\land {\sf Open}(c_k,d_k)$  \\
					$\land \pi(\phi(j))\neq \pi(\phi(k))$};
				
	\end{tikzpicture}}}
	\caption{The left figure is the interactive ZKP. The right figure is the non-interactive ZKP using Fiat-Shamir heuristic. In each protocol, the steps are listed in time order. In the non-interactive protocol shown in (b), $r=H(c)$ is taken as random. While in reality, the hash function $H$ is deterministic and cannot generate any randomness.}
	\label{fig:two}
\end{figure}
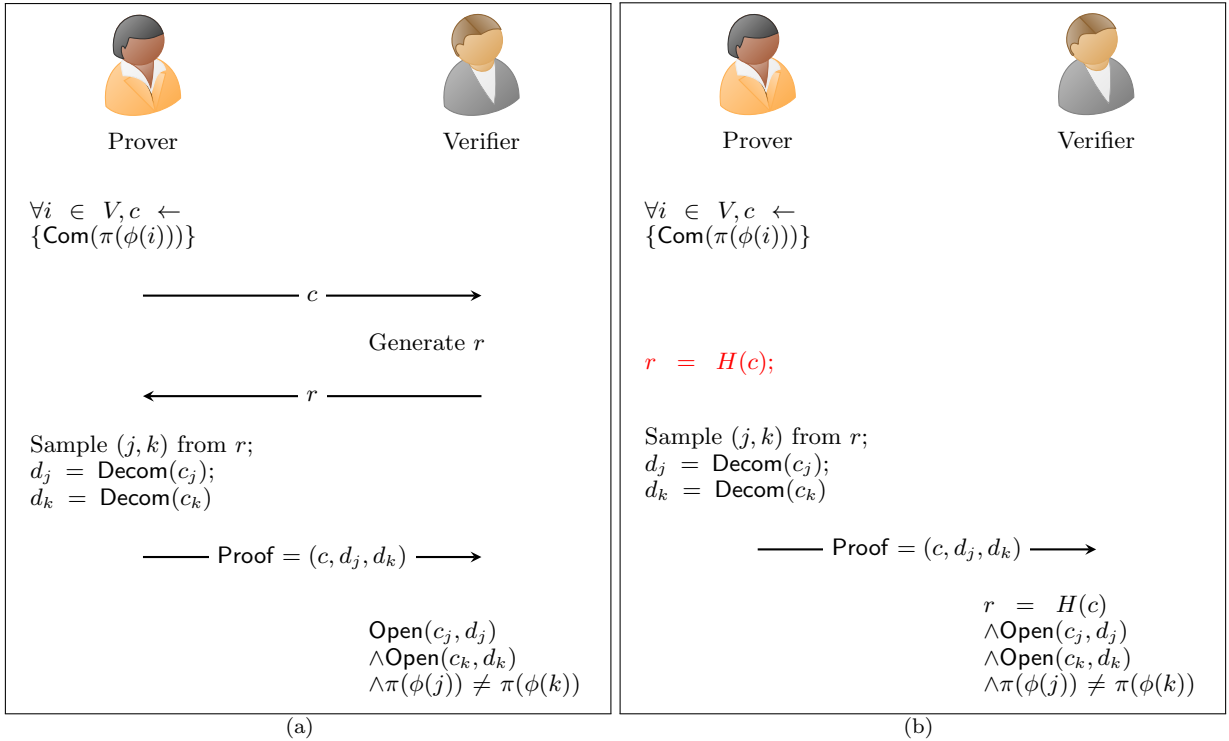

We briefly introduce the interactive ZKP and the non-interactive ZKP using Fiat-Shamir heuristic in Fig.~\ref{fig:two}. The origin interactive ZKP requires a truly uniform random number $r$. However, in the Fiat-Shamir heuristic, the prover calculates the hash of the commitment and  regard it as the random number.

\section{Post-Quantum-Cryptography Signature Algorithm}\label{sec:PQC}
In general, a lattice-based post-quantum-cryptography (PQC) signature is more complicated than its classical counterparts such as RSA and ECDSA. We briefly introduce our PQC digital signature algorithm, Aigis.Sig\cite{zhang2020tweaking}, which is based on the ``Fiat-Shamir with Aborts'' technique and can be seen as a variant of the NIST PQC round-3 finalist CRYSTALS-DILITHIUM~\cite{ducas2018crystals}.

\textit{Preliminary.}
Let $R_q=\mathbb{Z}[X]/(X^n+1)$ denote the quotient ring containing all polynomials over the $\mathbb{Z}_q$ in which $X^n$ is identified with $-1$.
Let ${\rm Hash}(\cdot)$ denote a hash function.
Let $||\cdot||_\infty$ denote the maximum norm. 
Let ${\rm HighBits}(r,\alpha)=\lfloor r/\alpha \rfloor$ and ${\rm LowBits}(r,\alpha)= r \bmod \alpha$ denote the higher-order and lower-order bits of $r$ with respect to the divisor $\alpha$, respectively.  $S_\eta$ denotes the set of ring elements of $R$, where each coefficient is taken from the set \{$-\eta$, $-\eta+1$, $\ldots$, $\eta$\} for some positive integer $\eta\ll q$.
Let $n,q,k,l,\eta,\gamma_1,\gamma_2,\beta$ denote other parameters. The PQC authentication algorithm includes the key generation, signature and verification algorithms, as given by Algorithms 1, 2, and 3, respectively. The procedure has at least 128-bit quantum security (against any quantum algorithms who attempt to forge a valid signature) based on the underlying quantum hardness of the lattice problems, and the correctness is ensured in the sense that any legitimate signature can be correctly verified by the verification algorithm.  
Finally, we remark that the ``repeat until'' subroutine in the signature algorithm represents the ``rejection sampling'' technique, which is necessary to sample from the desired distribution for security purposes and takes only up to a handful of trials.

\begin{algorithm}
\label{alg:gen}
\caption{Key Generation Algorithm}

\SetKwProg{keygen}{Function KeyGen}{}{end}

\keygen{}{
    $A\leftarrow R_q^{k\times l}$ \;
    $s_1,s_2\leftarrow S_{\eta}^l \times S_{\eta}^l$ \;
    $t=As_1+s_2$ \;
    $pk=(A,t),sk=(s_1,s_2,pk)$ \;
    \Return $(pk,sk)$  
}
\end{algorithm}

\normalem
\begin{algorithm}
\label{alg:decode}
\caption{Signature Algorithm}

\SetKwProg{sign}{Function Sign}{}{end}

\sign{$sk=(s_1,s_2,pk),\mu$}{

    \Repeat{$ ||z||_\infty < \gamma_1 - \beta$ and ${\rm LowBits}(Ay-cs_2,2\gamma_2)<\gamma_2-\beta$}
    {
        $y \leftarrow S_{\gamma_1-1}^{l+k}$ \;
        $w=Ay$ \;
        $c={\rm Hash}({\rm HighBits}(w,2\gamma_2) || \mu)$ \;
        $z=y+cs_1$ \;
    }
    \Return $\sigma=(z,c)$
}
\end{algorithm}
\ULforem

\normalem
\begin{algorithm}
\caption{Verification Algorithm}

\SetKwProg{verify}{Function Verify}{}{end}

\verify{pk=(A,t),$\sigma=(z,c),\mu$}{

    \If{$||z||_\infty<\gamma_1-\beta$ and $c={\rm Hash}({\rm HighBits}(Az-ct,2\gamma_2) || \mu)$}
    {
        \Return true \;
    }
    \Return false \;
}
\end{algorithm}
\ULforem

We implement the PQC algorithm in Windows 10 64-bit, Intel(R) Core(TM) i7-9750H CPU @2.60GHz, 8 GB RAM. The average CPU cycle of signature generation is 459903. The average CPU cycle of signature verification is 104337. The signature size is 2445 bytes. The real execution time is less than 1 ms.

\section{Randomness Service}\label{sec:randomness_service}


\subsection{Components}
As shown in Figure. 2(a) in the Main Text, our randomness beacon contains the following components,
\begin{enumerate}
    \item Device-independent random number generator (DIQRNG): A random number generator of true random numbers based on the quantum nonlocal behaviour, which requires no prior assumptions or characterisations of the inner working status of the quantum devices. 
    \item Security Module: A module containing cryptographic keys and performing cryptographic operations.
    \item Timestamp: A server synchronised with Coordinated Universal Time (UTC) promising to release the beacon pulse at the correct time. 
    \item Beacon App: The internal parts of the beacon service where the actual pulses are formed according to the pulse format in \ref{sec:pulse format} and uploaded to the web server, with the inputs from DIQRNG, timestamp and security module.
    \item Web Server: The public-facing parts of the beacon, storing beacon pulses in a database and providing a web interface to answer requests of information. All past pulses, and certain associated data, are stored in the database.
\end{enumerate}

The above components form the randomness beacon, the essence of our randomness service. 
In addition, we also provide a signature service, providing signed timestamps with PQC algorithms on the requirement. We remark that the timestamps for the beacon are independent of those for the signature service, where we block the beacon service from receiving information from the outside.

In our construction of the randomness service, compared to the existing counterparts, there are two key differences that lead our service to a higher security level. The first is that our Beacon integrates a DIQRNG based on the photonic loophole-free violation of a Bell inequality. The second is that a lattice-based AIGIS post-quantum-cryptography (PQC) signature algorithm is parallelly used with RSA4096 signature algorithm to identify the randomness server. 

Thanks to the experimental development, we are now able to establish an entropy source based on DIQRNG. Within the current understanding, this serves as the most secure means to generate true random numbers, providing intrinsic randomness and guaranteeing quantum security. Furthermore, we strive to increase the generation rate of DIQRNG, such that it meets the application requirements of random beacons. Moreover, in the establishment of our randomness beacon, we do not need multiple random number sources to avoid certain malicious behaviours, whereas the DIQRNG entropy source cannot be manipulated.

After the generation of random numbers, the post-processing security module is implemented by software, which indicates that both the RSA4096 algorithm and the AigisPQC algorithm are encapsulated into a program independent of the beacon APP. Limited by the current Internet public key infrastructure, RSA4096 signature and AigisPQC signature are both self-signed. In the first step, the two sets of public and private keys required for signature and verification are generated by two independent programs, RSA4096 and AigisPQC, where the required random numbers come from our DIQRNG. In the second step, the beacon APP calls the RSA4096 program and the Aigis program to sign and then generate the beacon pulse, so the private key used for signature is independent of the beacon APP. Relevant information, such as the public key and program required for the verification, are available on the Random Beacon website.

The above two improvements allow us to have a more secure, or, `quantum-safe' randomness beacon. It should be highlighted that in our randomness beacon, all hash functions are SHA512. Of course, the security property of a hash chain is also keeps valid here, which means that changing any record requires changing all future records.


\subsection{Pulse Format of the Randomness Beacon Service}\label{sec:pulse format}

In ~\cite{kelsey2018new,kelsey2019reference}, a new randomness beacon format standard has been formed to make the beacon easier to operate and more trustworthy as a third party.
The beacon app core implements beacon operations to generate beacon pulse according to the information entered, local random number from the entropy source, time-stamp from the time sever and signature algorithm from the security module. Our beacon pulse format is inherited from NIST Randomness Beacon version 2.0 but owning 25 fields \cite{kelsey2019reference}. All the fields and the corresponding value are listed as follows:
\begin{table}
    \centering
    \caption{Table 1. Field names, types, and interpretation}
    \begin{tabular}{|c|c|c|p{0.65\textwidth}<{\centering}|}
    \hline  \# & \text{Field name} & \text{Type} & \text{Interpretation} \\
    \hline $\mathrm{F}_{1}$ & \text{URL} &\text{Str} & \tabincell{c}{uniform resource locator that uniquely identifies the pulse} \\
    \hline $\mathrm{F}_{2}$ & \text{VERSION} &\text{Str} & \tabincell{c}{the version of the beacon format being used} \\
    \hline $\mathrm{F}_{3}$ & \text{CIPHER\_SUITE} & \text{uint64} & \tabincell{c}{the cipher suite (set of cryptographic algorithms) being used} \\
    \hline $\mathrm{F}_{4}$ & \text{PERIOD} &\text{uint64} & \tabincell{c}{the number of milliseconds between the timestamps of this pulse and the \\ expected subsequent pulse}\\
    \hline $\mathrm{F}_{5}$ & \text{CERTIFICATE\_ID} & \text{hashOut} & \tabincell{c}{the hash of the certificate that allows verifying the signature in the pulse} \\
    \hline $\mathrm{F}_{6}$ & \text{CHAIN\_INDEX} & \text{unit64} & \tabincell{c}{the chain index (integer identifier, starting at 1), of the chain to which the pulse \\ belongs} \\
    \hline $\mathrm{F}_{7}$ & \text{PULSE\_INDEX} & \text{uint64} & \tabincell{c}{the pulse index (integer identifier, starting at 1), conveying the order of \\ generation of this pulse within its chain} \\
    \hline $\mathrm{F}_{8}$ & TIMESTAMP & \text{Str} & \tabincell{c}{the time (UTC) of pulse release by the Beacon Engine} \\
    \hline $\mathrm{F}_{9}$ & LOCAL\_RANDOM\_VALUE & \text{hashOut} & \tabincell{c}{the hash() of high-quality random bit sources} \\
    \hline $\mathrm{F}_{10}$ & EXTERNAL\_SOURCE\_ID & \text{hashOut} & \tabincell{c}{the hash() of a text description of the source of the external value} \\
    \hline $\mathrm{F}_{11}$ & EXTERNAL\_STATUS\_CODE & \text{uint64} & \tabincell{c}{ the status of the external value} \\
    \hline $\mathrm{F}_{12}$ & EXTERNAL\_VALUE & \text{hashOut} & \tabincell{c}{the hash() of an external value} \\
    \hline $\mathrm{F}_{13}$ & PREVIOUS & \text{hashOut} & \tabincell{c}{the outputValue of the previous pulse} \\
    \hline $\mathrm{F}_{14}$ & HOUR & \text{hashOut} & \tabincell{c}{the outputValue of the first pulse in the (UTC) hour of the previous pulse} \\
    \hline $\mathrm{F}_{15}$ & DAY & \text{hashOut} & \tabincell{c}{the outputValue of the first pulse in the (UTC) day of the previous pulse} \\
    \hline $\mathrm{F}_{16}$ & MONTH & \text{hashOut} & \tabincell{c}{the outputValue of the first pulse in the (UTC) month of the previous pulse} \\
    \hline $\mathrm{F}_{17}$ & YEAR & \text{hashOut} & \tabincell{c}{the outputValue of the first pulse in the (UTC) year of the previous pulse} \\
    \hline $\mathrm{F}_{18}$ & PRECOMMITMENT\_VALUE & \text{hashOut} & \tabincell{c}{the hash() of the next pulse’s localRandomValue} \\
    \hline $\mathrm{F}_{19}$ & STATUS\_CODE & \text{uint64} & \tabincell{c}{the status of the chain at this pulse} \\
    \hline $\mathrm{F}_{20}$ & TYPE & \text{Str} & \tabincell{c}{the principal type of the random number source} \\
    \hline $\mathrm{F}_{21}$ & CHSH & \text{Str} & \tabincell{c}{the violation value of the CHSH inequality} \\
    \hline $\mathrm{F}_{22}$ & METHOD & \text{Str} & \tabincell{c}{the theoretical scheme of the evaluation method of the generated randomness \\ from the Bell test} \\
    \hline $\mathrm{F}_{23}$ & SIGNATURE\_RSA & \text{sigOut} & \tabincell{c}{a RSA signature on all the above 22 fields} \\
    \hline $\mathrm{F}_{24}$ & SIGNATURE\_PQC & \text{sigOut} & \tabincell{c}{a PQC signature on all the above 22 fields} \\
    \hline $\mathrm{F}_{25}$ & OUTPUT\_VALUE & \text{hashOut} & \tabincell{c}{the hash() of all the above fields} \\
    \hline
    \end{tabular}
    \label{tab:my_label}
\end{table}

A beacon pulse is a structure composed of 25 fields, of which: eleven (11) are hash outputs; two (2) are signature outputs; six (6) are characters strings; six (6) are unsigned integers.
The following is a brief introduction of partial fields and fields values. In our randomness beacon implementation, the URL for locating the pulse is
``https://www.randomnessbeacon.com/Shanghai/beacon/1.0/chain/1/pulse/N'' 
where N is the index. We can get all information of the pulse from the URL link. For our experiment, we get the pulse information from the database utilizing application programming interface (API) provided by randomness beacon website. The value of PERIOD field is 60000, that is to say the pulsating period of a pulse is one minute. The value of TIMESTAMP field refers to promised release time of this pulse by the beacon in the Universal Time Coordinated (UTC) standard. In our beacon, time is synchronized with a time server (www.tsa.cn) using  the network time protocol (NTP). The value of LOCAL\_RANDOM\_VALUE field is hash of random numbers generated from local RNG. High-quality random numbers are critical for the randomness beacon and it must be entirely unpredictable to any attacker. Usually, several RNGs are combined to bring security advantage and avoid some malicious RNGs. The five fields, PREVIOUS, HOUR, DAY, MONTH and YEAR, are records of OUTPUT values pertaining to some past pulses, which can ensure efficient hash-chain verification to prove that an old pulse is consistent with a recent pulse.
The value of PRECOMMITMENT\_VALUE field is the hash of LOCAL\_RANDOM\_VALUE field value of the next beacon pulse, so the PRECOMMITMENT\_VALUE and LOCAL\_RANDOM\_VALUE fields can be used together to combine several beacons and achieve security against misbehavior by partial beacons in future randomness beacon networks. The SIGNATURE\_RSA and SIGNATURE\_PQC fields are independent signature of the hash() of a concatenation of previous fields from $\mathrm{F}_{1}$ to $\mathrm{F}_{22}$. The OUTPUT value is the hash() of a concatenation of all previous fields from $\mathrm{F}_{1}$ to $\mathrm{F}_{24}$.

Besides, three special fields are added to the beacon pulse, TYPE, CHSH and METHOD: TYPE means the principal type of the random number source, the value of the TYPE field in our randomness beacon is device-independent quantum random number generation (DIQRNG); CHSH means the violation value of the CHSH inequality, the value of the CHSH field in our randomness beacon is about 2.007; METHOD means the theoretical scheme of the evaluation method of the generated randomness from the Bell test, the value of the MRTHOD field in our randomness beacon is quantum probability estimation (QPE) which has been explained in section~\ref{sec:experiment}.

\section{Device-Independent Quantum Random Number Generator}
The device-independent quantum random number generator (DIQRNG) is the core component of the randomness beacon. Here, we introduce the details of the theory and experiment of our DIQRNG.

\subsection{Theory of Device-Independent Quantum Random Number Generator}\label{theory}
To achieve the highest level of security and reach the requirement of unpredictability, we adopt the DIQRNG for the beacon service. In general, a DIQRNG is composed of a device-independent quantum randomness generation process based on the Bell test and a randomness extraction procedure. In the Bell test, the generated raw data contains freshly generated randomness, while the probability distribution of the raw data may be highly biased. After the amount of randomness is evaluated, the raw data is post-processed with randomness extraction. At the end of this process, near-uniformly distributed random numbers shall be output.

In this section, we first introduce the theoretical basis of randomness generation and randomness extraction. Afterwards, we present the detailed protocol description for DIQRNG and the rigorous security definitions.

\subsubsection{Randomness Evaluation via Quantum Probability Estimation}\label{sec:RandEval}
For randomness generation, we adopt the Clauser-Horne-Shimony-Holt-type (CHSH) Bell test. The configuration involves two remote stations, commonly referred to as Alice and Bob. In one round of the CHSH test, the two parties take random measurements with untrusted devices. Alice and Bob each randomly set the measurement inputs, denoted by random variables $X,Y \in \{0,1\}$, respectively. Their devices shall generate binary outcomes, which form another two random variables $A,B$ ranging in the set $\{0,1\}$. We use the lowercase letters to represent specific realisations of the associating random variables. 

In the randomness generation procedure, we carry out loophole-free Bell tests sequentially. By `sequential', we refer to that only after the devices have generated outputs shall the inputs in the subsequent round be fed into the devices. To distinguish different rounds, we use subscripts to denote the associating random variables. In a loophole-free Bell test, the following assumptions or requirements are made,
\begin{enumerate}
    \item \emph{Quantum mechanism:} Quantum mechanics is correct and complete.
    \item \emph{Non-signaling:} In each round, the measurement process of Alice/Bob is independent of the other party.
    \item \emph{Trusted inputs:} Alice and Bob each have trusted random input settings.
\end{enumerate}
To meet the non-signalling assumption, we pose a space-like separated configuration between Alice's and Bob's devices. To meet the trusted-inputs assumption, we use two local quantum random number generators for the input settings. Under these assumptions, if the untrusted devices have a nonlocal behaviour, intrinsic randomness shall be generated from the outputs. The nonlocal behaviour can be witnessed by quantum inequalities. To give an intuitive explanation, take CHSH Bell inequality as an example. In the CHSH Bell test, we say the two parties win the game if
\begin{equation}
    A\oplus B=X\cdot Y.
\end{equation}
If the untrusted devices are restricted to classical resources, the winning probability is upper bounded,
\begin{equation}
    \omega_{CHSH}:=\Pr(A\oplus B=X\cdot Y)\leq\frac{3}{4}.
\end{equation}
The inequality is called CHSH inequality, and the bound is called the classical bound. If this is the case in an experiment, no intrinsic randomness can be generated. While if the untrusted devices utilise nonlocal quantum resources, the winning probability can reach at most $\frac{2+\sqrt{2}}{4}$, which is called the Tsirelson Bound. In evaluating the experimental behaviour, we shall use the CHSH value, which is linked with the winning probability via a linear transformation, 
\begin{equation}
  S=8\omega_{CHSH}-4.
\end{equation}
By the $n$th trial, the real-time CHSH violation value is defined as
\begin{equation}
    \bar{S}_n=8\left(\frac{\sum_{i=1}^n\chi(A_i\oplus B_i=X_i\cdot Y_i)}{n}\right)-4,
\end{equation}
where $\chi(\cdot)$ is the indicator function. 

In an experiment, if the observed winning frequency is above the classical bound, we say the untrusted devices have a nonlocal behaviour. In this case, the devices must be taking quantum measurements on an entangled state. From the monogamy property of entanglement, the state should have an independent property from any outside observer, hence the measurement results should contain private randomness from the outside.

To rigorously evaluate the amount of generated randomness from the Bell test, we apply the quantum probability estimation (QPE) method~\cite{knill2018quantum}. We summarise the essential results of the QPE theory and information-theoretic randomness measures. In the QPE framework, the first step is a characterisation of the set of possible final states in the experiment, which we call the model of the experiment. In our device-independent quantum randomness generation, in general, Alice and Bob and a potential adversary, Eve, share a tripartite quantum state $\rho_{ABE}\in\mathcal{D}(\mathcal{H}_{ABE})$ in the beginning. After Alice and Bob measure their own systems in the experiment, the possible final state can be described by a classical-quantum mixed state
\begin{equation}
    \rho_{\bm{XYAB}E}=\sum_{\bm{x},\bm{y},\bm{a},\bm{b}}\ket{\bm{xy}\bm{ab}}\bra{\bm{xy}\bm{ab}}\otimes\rho_E(\bm{xy}\bm{ab}),
\end{equation}
where we use bold letters to denote the sequences of values over the rounds, e.g., $\bm{x}=(x_1,\cdots,x_n)$, and $\rho_E=\Tr_{AB}(\rho_{ABE})$ corresponds to the quantum side information of Eve. Under our assumptions, in the $i$th round,  conditioned on the occured events $H_i\equiv(X_1^{i-1},Y_1^{i-1},A_1^{i-1},B_1^{i-1})=h_i\equiv(x_1^{i-1},y_1^{i-1},a_1^{i-1},b_1^{i-1})$, there exist measurements described by positive operator-valued measures (POVM) $\{\hat{N}_{x_i}^{a_i}\}_{a_i},\{\hat{N}_{y_i}^{b_i}\}_{b_i}$ corresponding to the input settings $X_i=x_i,Y_i=y_i$, where the positive semi-definite operator $\hat{N}_{x_i}^{a_i}$ denotes the operator corresponding to the input $x_i$ and output $a_i$ for Alice, and similarly for $\hat{N}_{y_i}^{b_i}$ on Bob's side, and $\sum_{a_i}\hat{N}_{x_i}^{a_i}=\hat{I}_A, \sum_{b_i}\hat{N}_{y_i}^{b_i}=\hat{I}_B$, and a quantum state $\rho_{x_1^{i-1}y_1^{i-1}a_1^{i-1}b_1^{i-1}E}$, such that
\begin{equation}
\begin{aligned}
    \Pr_{A_iB_i|X_iY_iH_i}(a_ib_i|x_iy_ih_i)=\Tr_{AB}\left[\rho_{x_1^{i-1}y_1^{i-1}a_1^{i-1}b_1^{i-1}E}(\hat{N}_{x_i}^{a_i}\otimes \hat{N}_{y_i}^{b_i}\otimes \hat{I}_E)\right].
\end{aligned}
\end{equation}
We denote the set of all possible final states shared by Alice, Bob, Eve after the experiment as the model $\mathcal{M}(\bm{X},\bm{Y},\bm{A},\bm{B})$. For the detailed description of the model, we refer the readers to our previous work in~\cite{li2021experimental}. 

For a model $\mathcal{M}(\bm{X},\bm{Y},\bm{A},\bm{B})$, we can define quantum estimation factors (QEF).
\begin{definition}[Quantum Estimation Factor] Given $\alpha>1$ and a model $\mathcal{M}(\bm{X},\bm{Y},\bm{A},\bm{B})$, the non-negative real-valued function $F(\bm{XYAB})$ is a quantum estimation factor (QEF) with power $\alpha$ for $\mathcal{M}(\bm{X},\bm{Y},\bm{A},\bm{B})$, if for any state $\tau\in\mathcal{M}(\bm{X},\bm{Y},\bm{A},\bm{B})$, $F(\bm{XYAB})$ satisfies the following inequality
    \begin{equation}
    \begin{aligned}
      &\sum_{\bm{x,y,a,b}}F(\bm{xyab})\mathcal{R}_{\alpha}[\tau_E(\bm{xyab})|\tau_E(\bm{xy})]\leq 1,\alpha > 1,
    \end{aligned}
    \end{equation}
where
\begin{equation}
  \tau = \sum_{\bm{x,y,a,b}}\ket{\bm{xyab}}\bra{\bm{xyab}}\otimes\tau_E(\bm{xyab}),\tau_E(\bm{xy}) = \sum_{\bm{ab}}\tau_E(\bm{xyab}),
\end{equation}
and the function
\begin{equation}
      \Renyi(\rho|\sigma) = \Tr\left[\left(\sigma^{-\beta/(2\alpha)}\rho\sigma^{-\beta/(2\alpha)}\right)^\alpha\right]
\end{equation}
is called the R\'enyi power of order $\alpha$ of $\rho$ conditional on $\sigma$.
\end{definition}

We can give a lower bound on the amount of randomness that is generated from the Bell test via QEFs, which is measured by the smooth conditional min-entropy~\cite{Tomamichel2010duality}.
\begin{definition}[Smooth Conditional Min-Entropy]
  Consider a quantum state $\rho\in\mathcal{D}(\mathcal{H}_{AE})$. The $\varepsilon$-smooth min-entropy of system $A$ conditioned on $E$ is
  \begin{equation}
  \begin{aligned}
    &\Hmin^{\varepsilon}(A|E)_{\rho} = \max_{\substack{P(\rho',\rho)\leq\varepsilon,\\ \rho'\in\mathcal{S}(\mathcal{H}_{ABE})}} \Hmin(A|E)_{\rho'},\\
    &\Hmin(A|E)_{\rho'} = \sup_{\sigma\in\mathcal{S}(\mathcal{H}_{E})}\sup_{\lambda}\{\lambda\in\mathbb{R}: \rho'\leq\exp(-\lambda)I_A\otimes\sigma\},
  \end{aligned}
  \end{equation}
where $P(\rho',\rho)$ is the purified distance between $\rho',\rho$, and $\mathcal{S}(\cdot)$ denotes the set of sub-normalised density operators acting on the corresponding Hilbert space. The purified distance between $\rho,\tau\in\mathcal{S}(\mathcal{H})$ is defined as
\begin{equation}
\begin{aligned}
  P(\rho,\tau)=\sqrt{1-\left(\Tr|\sqrt{\rho}\sqrt{\tau}|+\sqrt{(1-\Tr[\rho])(1-\Tr[\tau])}\right)^2}.
\end{aligned}
\end{equation}
\end{definition}

\begin{theorem}(Theorem 3 in~\cite{knill2018quantum})
Prior to a randomness generation procedure with $N$ trials, suppose that $F(\bm{XYAB})$ is a QEF with power $\alpha$ for the model of the experiment $\mathcal{M}(\bm{X},\bm{Y},\bm{A},\bm{B})$, and set the smoothing parameter in randomness generation $\varepsilon_h\in(0,1]$, the threshold $h_s>0$ for a successful randomness generation and a lower bound $\kappa\in(0,1]$ to the probability of a successful randomness generation.A successful randomness generation is that, after $N$ trials have been executed, an event $(\bm{x},\bm{y},\bm{a},\bm{b})$ happens such that $F(\bm{xyab})\geq 2^{h_s(\alpha-1)}$. In this case, for any possible final state $\rho\in\mathcal{M}(\bm{X},\bm{Y},\bm{A},\bm{B})$, we have \emph{either} the probability of success is less than $\kappa$, \emph{or} the quantum smooth conditional min-entropy given success satisfies
	\begin{equation}\label{eq:QEFRandomness}
	\Hmin^{\varepsilon_h}(\bm{XY}|\bm{AB} E)_{\rho}\geq h_s-\dfrac{1}{\alpha-1}\log_2\left(\dfrac{2}{\varepsilon_h^2}\right)+\dfrac{\alpha}{\alpha-1}\log_2\kappa.
	\end{equation}
\label{Thm:QPE}
\end{theorem}

\subsubsection{Randomness Extraction}\label{extraction}
The raw data from the Bell test cannot be directly used as the random numbers for the randomness beacon service, as the statistics is not uniformly distributed. A randomness extraction procedure needs to be performed, processing the raw data to be uniformly distributed while maintaining enough entropy. In general, the randomness extraction is a classical post-processing procedure. We take the following assumptions for the classical module in DIQRNG:
\begin{enumerate}
\item \emph{Secure lab: }The devices cannot communicate to the outside to leak the experimental results directly.
\item \emph{Trusted post-processors: }The classical post-processing procedure is trusted.
\end{enumerate}
The first assumption is also posed for the management of the raw data from the randomness generation procedure. 

To guarantee the randomness extraction output is information-theoretic random and uniformly distributed, we need to apply a quantum-proof strong extractor. Its definition is given as follows.
\begin{definition}[Quantum-Proof Strong Extractor~\cite{nisan1996randomness,konig2011sampling,de2012trevisan,ma2013postprocessing}] A function $\text{Ext: }\{0,1\}^n\times\{0,1\}^d\rightarrow\{0,1\}^m$ is a quantum-proof $(k,\varepsilon_x)$-strong extractor with a uniform seed, if for all classical-quantum states $\rho_{XE}$ classical on $X$ with $H_{\text{min}}(X|E)_\rho\geq k$ and a uniform seed $Y$, we have
\begin{equation}
    \dfrac{1}{2}\|\rho_{\text{Ext}(X,Y)YE}-\rho_{U_m}\otimes\rho_Y\otimes\rho_E\|_1\leq\varepsilon_x,
\end{equation}
where $\|\cdot\|_1$ is the trace norm defined as $\|A\|_1=\Tr(\sqrt{A^\dag A})$, $\rho_{U_m}=\frac{I}{2^m}$ is the fully mixed state of dimension $2^m$ and $\rho_Y$ is the fully mixed state of the seed $Y$.
\end{definition}

The definition states that, given a quantum-proof $(k,\varepsilon_x)$-strong extractor with a uniform seed, $m$ bits of uniformly distributed random bits can be provided (except for a failure probability no larger than $\varepsilon_x$) if there is a guarantee of $k$ bits of min-entropy in the input $X$.
For brevity, we call $\varepsilon_x$ the security parameter of the extractor. This definition naturally guarantees a composability property, shown by the following lemma:
\begin{lemma}[\cite{de2012trevisan,renner2008security}]
  If $\text{Ext: }\{0,1\}^n\times\{0,1\}^d\rightarrow\{0,1\}^m$ is a quantum-proof $(k,\varepsilon_x)$-strong extractor, then for any classical-quantum state $\rho_{XE}$ such that $H_{\text{min}}^{\varepsilon_h}(X|E)_\rho\geq k$, we have
\begin{equation}\label{Eq:Composable}
    \dfrac{1}{2}\|\rho_{\text{Ext}(X,Y)YE}-\rho_{U_m}\otimes\rho_Y\otimes\rho_E\|_1\leq\varepsilon_x+2\varepsilon_h.
\end{equation}
\end{lemma}

From this property, we can reuse the extraction seed albeit a linear increase in the security parameter.

In our experiment, we use the Toeplitz extractor in the randomness extraction. An $m\times n$ Toeplitz matrix has $(m+n-1)$ bits of free parameters, taking the form
\begin{equation}
  T =
  \left(
    \begin{array}{ccccc}
    a_0     & a_{-1}  & \cdots & a_{-(n-2)}   & a_{-(n-1)} \\
    a_1     & a_0     & \ddots &              & a_{-(n-1)+1} \\
    a_2     & a_1     & \ddots & \ddots       & \vdots     \\
    \vdots  & \vdots  &        & \ddots       & a_{-(n-1)+(m-2)} \\
    a_{m-1} & a_{m-2} & \cdots & a_{-n+(m-1)} & a_{-(n-1)+(m-1)} \\
    \end{array}
  \right),
\end{equation}
where $\forall i-j=i'-j', T_{ij}=T_{i'j'}\in\{0,1\}$. For a family of $m\times n$ Toeplitz matrices, we have the following lemma that guarantees its use as a quantum-proof strong extractor:
\begin{lemma}[\cite{renner2008security}]
  The set of all $m\times n$ Toeplitz matrices can be used as a quantum-proof $(k,\varepsilon_x)$-strong  extractor with $\varepsilon_x=2^{-(k-m)/2}$, where $m$ is the output length and $n$ is the input data length.
\end{lemma}
The Toeplitz extractor is a quantum-proof strong extractor~\cite{Impagliazzo:Leftover:1989, frauchiger2013true, ma2013postprocessing}. In applying the Toeplitz extractor, one can adopt the fast Fourier transformation method to accelerate the computation. We refer the readers to the Supplementary Information of~\cite{liu2018device} for a detailed description of the processing.

\subsubsection{Protocol and Security Definition}
In this subsection, we shall discuss the security of the protocol under the QPE framework. We adopt the composable security definition~\cite{canetti2000security,ben2004general,portmann2014cryptographic}:
\begin{tcolorbox}[title = {Box 1. DIQRNG Protocol}]
\textbf{Input: }

$k_{\text{gen}}$: number of random bits to be generated \\
$\varepsilon_S$: soundness error \\
$S$: uniform seed of length $d$\\

\textbf{Given: }\\
$N$: number of rounds in randomness generation \\
$\mathcal{M}(\bm{X},\bm{Y},\bm{A},\bm{B})$: model of the experiment \\
$F(\bm{XYAB})$: valid QEF with power $\alpha$ for $\mathcal{M}(\bm{X},\bm{Y},\bm{A},\bm{B})$ \\
$\text{Ext}(,)$: quantum-proof $(k,\varepsilon_x)$-strong extractor, a map $\{0,1\}^{2N}\times\{0,1\}^d\rightarrow\{0,1\}^{k_{\text{gen}}}$ \\

\textbf{Output: }\\
$U_{k_{\text{gen}}}$: bit string of length $k_{\text{gen}}$

\tcblower
\textbf{Parameter Assignments}
\begin{enumerate}
\item Assign the smoothing parameter in randomness generation $\varepsilon_h$, the security parameter of the extractor
    $\varepsilon_x$ such that $\varepsilon_S = 2\varepsilon_h + \varepsilon_x$.
\item Assign the success threshold of the randomness generation $h_s$ and a lower bound on the protocol success probability $\kappa$.
\end{enumerate}
\textbf{Randomness Generation}
\begin{enumerate}
\setcounter{enumi}{2}
\item Perform the randomness generation experiment and obtain a realisation $(\bm{x},\bm{y},\bm{a},\bm{b})$ of $(\bm{X},\bm{Y},\bm{A},\bm{B})$ and the QEF value $F(\bm{xyab})$.
\item If $F(\bm{xyab})\geq2^{h_s}(\alpha-1)$, the protocol proceeds to randomness extraction; otherwise, the protocol fails and return $U_{k_{\text{gen}}} = 0^{\frown k_{\text{gen}}}$.
\end{enumerate}
\textbf{Randomness Extraction}
\begin{enumerate}
\setcounter{enumi}{4}
\item Obtain a realisation $s$ of the uniform seed $S$ of length $d$.
\item Return $U_{k_{\text{gen}}} = \text{Ext}(\bm{c},s)$.
\end{enumerate}

\end{tcolorbox}
\begin{definition}[Protocol soundness and completeness]
  A random number generation protocol with an $m$-bit output $\bm{C}$ is called $(\varepsilon_S,\varepsilon_C)$-secure if it is\\
  (1) $\varepsilon_S$-Sound: The output satisfies
  \begin{equation}
    \dfrac{1}{2}p_{\Phi}\|\rho_{\bm{C}E|\Phi} - \tau_m\otimes\rho_{E|\Phi}\|_1\leq\varepsilon_S,
  \end{equation}
  where $\Phi$ represents the event that the protocol does not abort, $\rho_{\bm{C}E|\Phi}$ is the normalised final state conditioned on a success, with $\rho_{\bm{C}}$ giving the output $\bm{C}$, and $\rho_E$ the system of side information, and $\tau_m$ is a maximally mixed state of $m$ qubits. The probability that the protocol does not abort is $p_{\Phi}$. \\
  (2) $\varepsilon_C$-Complete: There exists an honest implementation where the protocol
  observes an identical and independent behaviour, and the probability that it does not abort satisfies $p_{\Phi}\geq 1-\varepsilon_C$.
\end{definition}

In our experiment, we apply the same random number generation protocol as in~\cite{li2021experimental} (Appendix A, Box 1).

Under the QPE framework, the soundness and completeness for the protocol have been proved (see Sec. IIIC of~\cite{knill2018quantum}).
The total soundness error of the protocol is $\varepsilon_S = \varepsilon_x + 2\varepsilon_h$ as given in Eq.~\eqref{Eq:Composable}.

\subsubsection{Parameter Determination}\label{sec:ParDet}
Before randomness generation, we need to determine the largest allowed number of experimental trials, the soundness error and the failure probability of the protocol, and find a valid QEF that yields a good randomness generation rate. The determination procedure is similar to the one in our previous work of device-independent quantum randomness expansion (see Appendix B of~\cite{li2021experimental}). Here, we summarise the differences between the two tasks:

\begin{enumerate}
\item The randomness expansion experiment in~\cite{li2021experimental} takes a spot-checking protocol, which involves an additional coordinating random number generator to determine if a trial is ``spot'', where both detection stations take fixed inputs, or ``checking'', where the detection stations carry out a standard Bell test with uniformly local random inputs. In the randomness generation process of this work, the experimental set-up is the same as a standard Bell test, where both stations take uniformly random inputs. There is not an additional coordinating random number generator.
\item In this work, we do not consider the consumption of randomness in the training trials and the input setting in randomness generation. 
\item In randomness expansion, one needs to determine a proper spot-checking probability distribution that supports a net increase in the store of randomness.
\end{enumerate}

\begin{tcolorbox}[title = {Box 2. Parameter Determination Procedure}]
\begin{enumerate}
\item Set the input probability distribution $\nu(XY)$ as uniformly distributed. Conduct ``training trials'' and determine an empirical input-output probability distribution $\nu(ABXY)$ for randomness generation.
\item Determine a QEF $F(ABXY)$ with power $\alpha$ that yields a large $r_\nu (F,\alpha)$.
\item Set the randomness generation target $k$, the largest allowed number of trials $N$, the soundness error in randomness generation $\varepsilon_h$, the soundness error in randomness extraction $\varepsilon_x$, the completeness error of the protocol $\varepsilon_C$, and a lower bound on the success probability $\kappa$ of the protocol.
\item Set the success threshold in the randomness generation experiment $h_s$.
\end{enumerate}

\end{tcolorbox}

For the generation task to generate at least $k_{exp}$ bits of near-uniform random bits at the end of the protocol with failure probability $\varepsilon_x$ in randomness extraction, the amount of $\varepsilon_h$-smooth min entropy that needs to be generated $k$ is determined by 
\begin{equation}
    k = k_{exp} - 2\log_2\varepsilon_x.
\end{equation}
According to Theorem~\ref{Thm:QPE}, the $\varepsilon_h$-smooth min entropy is linked with the threshold $h_s$ in the QPE-based protocol,
\begin{equation}
    h_s=k+\frac{1}{\alpha-1}\log_2\left(\frac{2}{\varepsilon_h^2}\right)+\frac{\alpha}{\alpha-1}\log_2\left(\frac{1}{\kappa}\right),
\end{equation}
where $\kappa$ is a pre-determined successful probability of the protocol.

\subsection{Experiment of Device-Independent Quantum Randomness Generation}\label{sec:experiment}

\subsubsection{System Characterization}

\begin{figure}
    \centering
    \includegraphics[width = 0.7\textwidth]{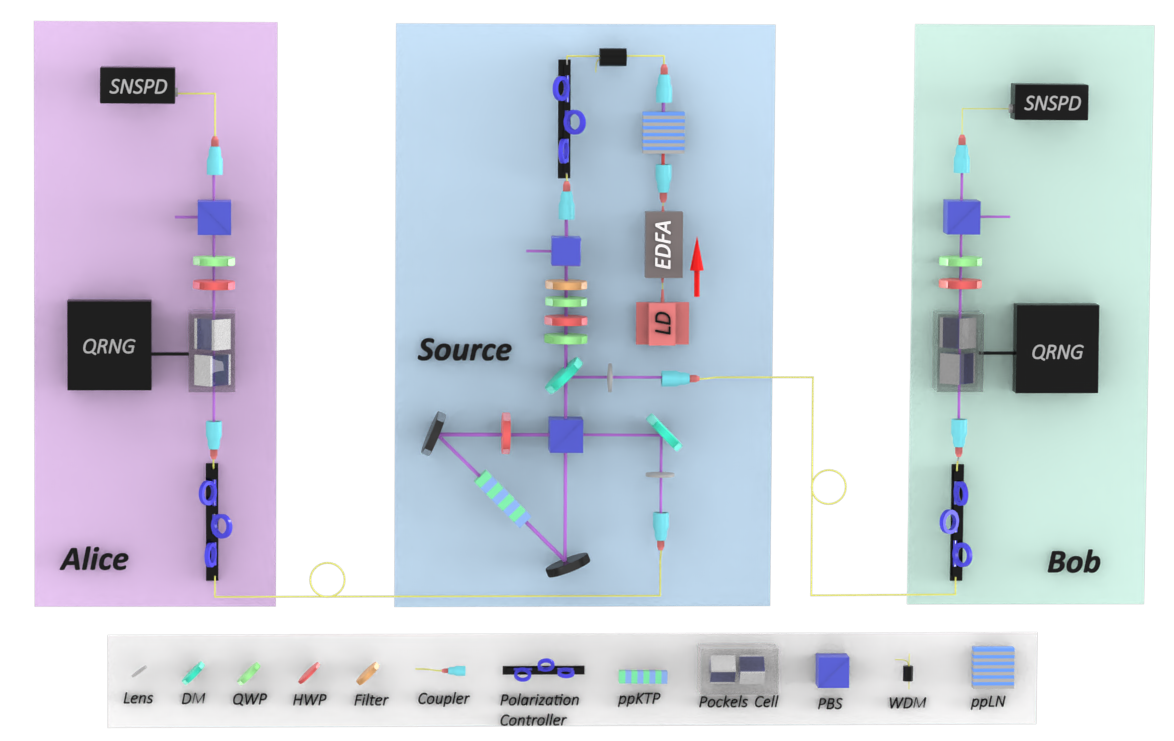}
    \caption{ Light pulses of width of 10ns and wavelength of 1560nm at the repetition frequency of 200kHz from laser diode (LD) are amplified by erbium-doped fibre amplifier (EDFA) and frequency-doubled by periodically poled lithium niobate crystal (PPLN) as a pump beam. The pump pulses are injected into a Sagnac loop to generate ploarization entangled photon pairs at a  periodically poled potassium titanyl phosphate (PPKTP) crystal. We use a half-wave plate (HWP) and two quarter-wave plate (QWP) to adjust the relative amplitude and phase of the entangled two-photon state. The two detection stations (denoted as Alice and Bob) are placed on the opposite directions from the entangled-photon source. Using the nonlinearity of the Pockels cell, we change the voltage applied on the Pockels cell according to the input random numbers from a quantum random number generator (QRNG), thereby randomly selecting the basis vector for measurement. The measurement results are detected by a superconducting nanowire single photon detector (SNSPD). DM, dichroic mirror; WDM, wavelength-division multiplexer; PBS, polarizing beam splitter.
    }
    \label{fig:DI_setup_v15}
\end{figure}
As shown in Fig. \ref{fig:DI_setup_v15}, we perform the protocol on our photonic-entanglement distribution platform \cite{li2018test,liu2018high,liu2018device,li2021experimental}. 

\subsubsection{Determination of single photon efficiency}
In Table \ref{tab:efficiency} we list the heralding efficiencies from the entanglement source to detection stations, the transmission efficiencies of each intermediate process, and the detection efficiencies of the detectors. We determine the single photon heralding efficiency as $\eta_A = C/N_B$ and $\eta_B = C/N_A$ for Alice and Bob, in which two-photon coincidence events $C$ and single photon detection events for Alice $N_A$ and Bob $N_B$ are measured in the experiment. We measure the transmission efficiencies of each process and the detection efficiencies with classical light beams and NIST-traceable power metres, where $\eta^{sc}$ is the efficiency to couple entangled photons into single mode optical fibre, $\eta^{so}$ is the efficiency for photons passing through the optical elements in the source, $\eta^{fibre}$ is the transmittance of fibre connecting source to measurement station, $\eta^m$ is the efficiency for light passing through the measurement station, and $\eta^{det}$ is the single photon detector efficiency.

\begin{table}
    \centering
    \caption{Photon transmission and detection efficiencies in the experiment.}
    \begin{tabular}{c c c c c c c}
    \hline
    Parties & Heralding,$\eta$ & $\eta^{sc}$ & $\eta^{so}$ & $\eta^{fibre}$ & $\eta^{m}$ & $\eta^{det}$\\
    \hline
    Alice & 80.55\% & 93.5\% & \multirow{2}{*}{95.9\%} & \multirow{2}{*}{99\%} & 94.5\% & 96.0\% \\
    Bob & 82.13\% & 93.6\% & ~ & ~ & 95.2\% & 97.1\% \\
    \hline
    \end{tabular}
    \label{tab:efficiency}
\end{table}

\subsubsection{Quantum state and measurement bases}

\begin{figure}
    \centering
    \subfigure[]{
        \includegraphics[width = 0.45\textwidth]{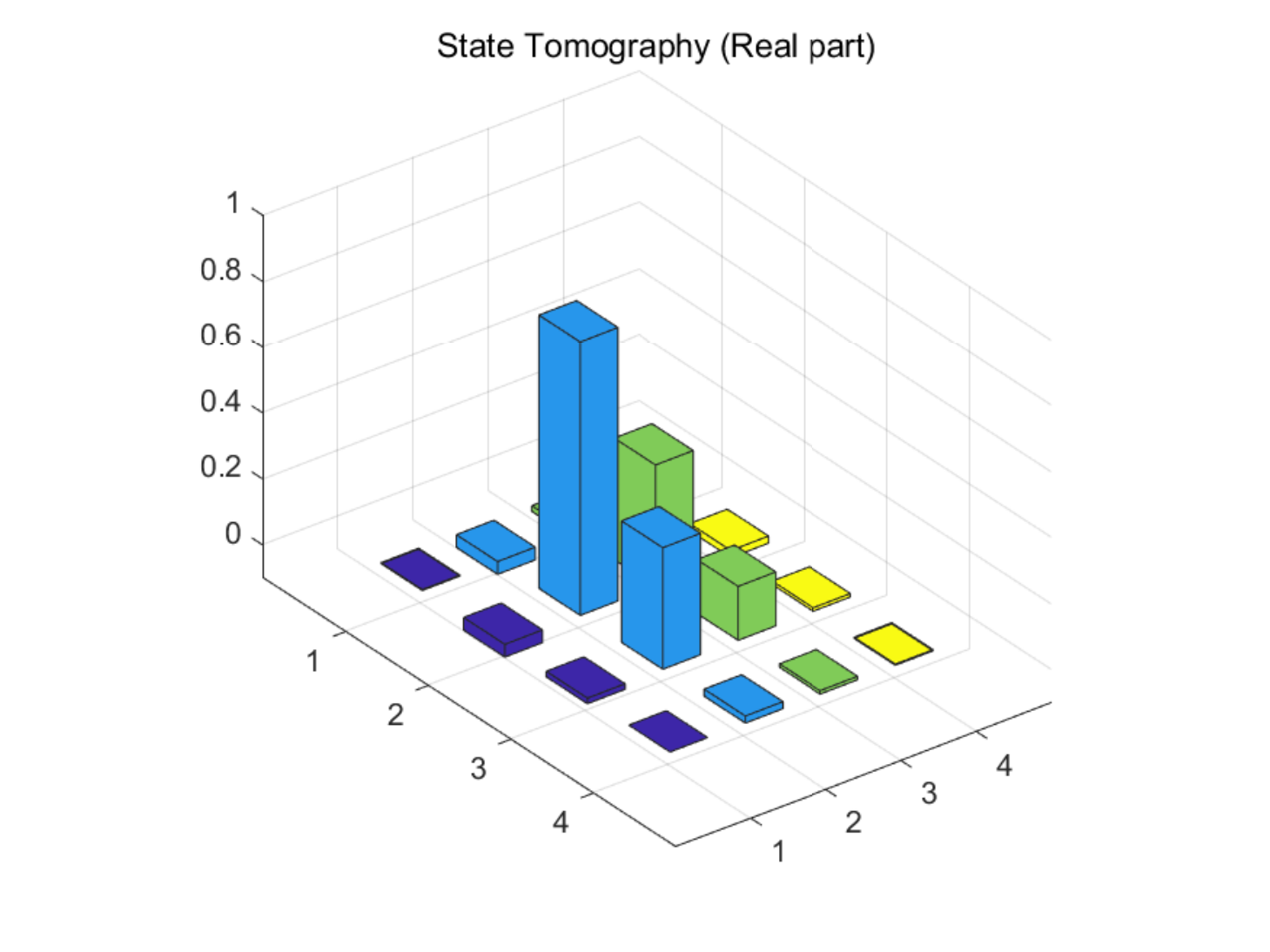}
    }
    \subfigure[]{
    \includegraphics[width = 0.45\textwidth]{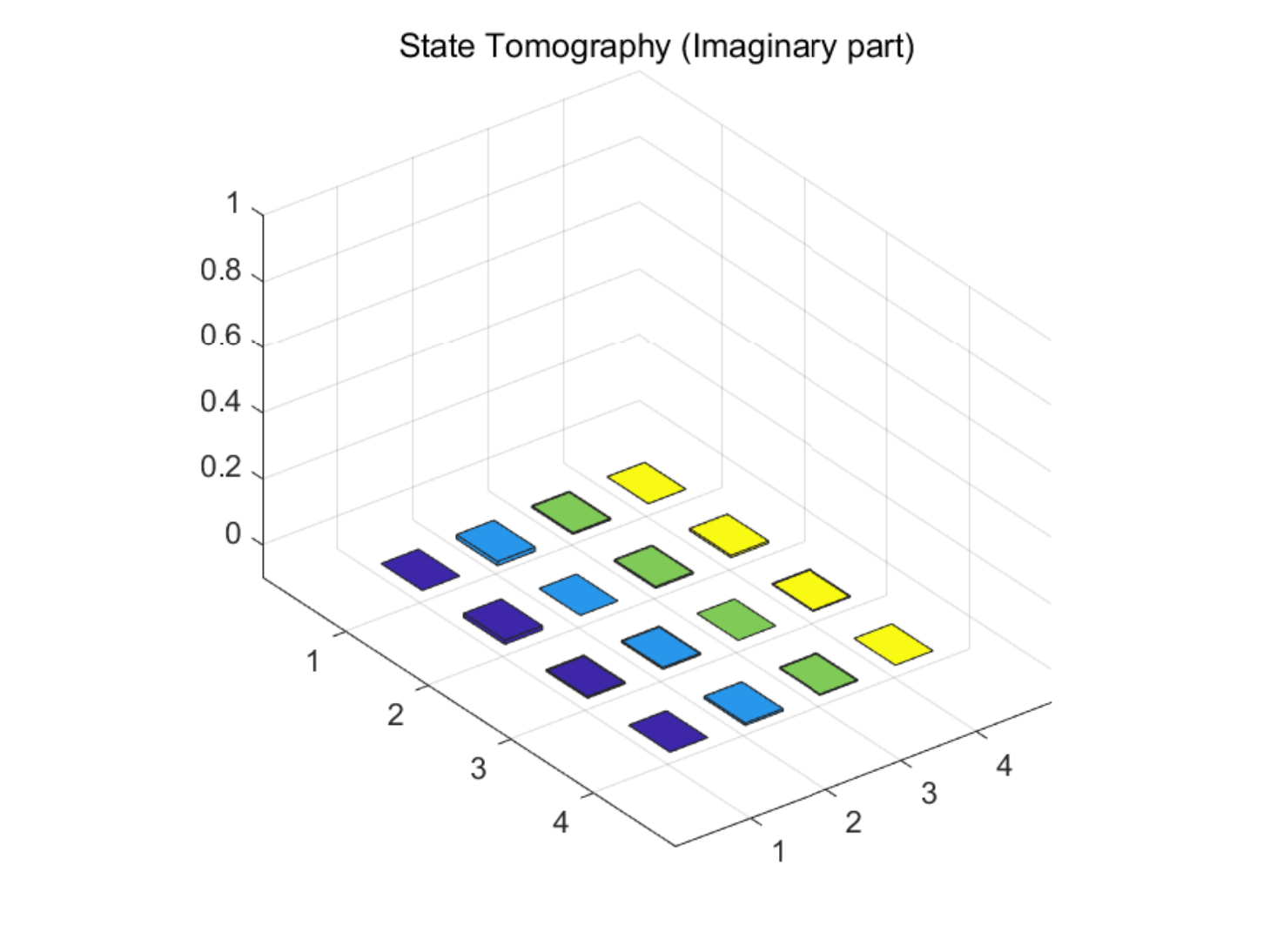}
    }
    \caption{Tomography of the produced two-photon state in the experiment, with real and imaginary components shown in (a) and (b), respectively.}
    \label{fig:tomo}
\end{figure}

To maximally violate the Bell inequality in experiment, we create non-maximally entangled two-photon state \cite{Eberhard93}
$\cos(24.39^{\circ})\ket{HV}+\sin(24.39^{\circ})\ket{VH}$ and set measurement bases to be $A_1= -83.05^{\circ}$ and $A_2=-118.64^{\circ}$ for Alice,
and $B_1=6.95^{\circ}$ and $B_2=-28.64^{\circ}$ for Bob, respectively.
We measure diagonal/anti-diagonal visibility in the bases set $(45^{\circ}, -24.39^{\circ})$, $(114.39^{\circ}, 45^{\circ})$ for minimum coincidence,
and in the bases set $(45^{\circ}, 65.61^{\circ})$, $(24.39^{\circ}, 45^{\circ})$ for maximum coincidence, where the angles represent measurement
basis $\cos(\theta)\ket{H}+\sin(\theta)\ket{V}$ for Alice and Bob. By setting the mean photon number to $\mu=0.266$ to suppress the
multi-photon effect, we measure the visibility to be 99.5\% and 98.4\% in horizontal/vertical basis and diagonal/anti-diagonal basis.

We perform quantum state tomography measurement of the non-maximally entangled state, with result shown in Fig.~\ref{fig:tomo}. The state fidelity reaches 99.33\%. We attribute the imperfection to multi-photon components, imperfect optical elements, and imperfect spatial/spectral mode matching.

\subsubsection{Spacetime configuration of the experiment}\label{sec:spacetime_details}
\begin{figure}
    \centering
    \includegraphics[width = 0.6\textwidth]{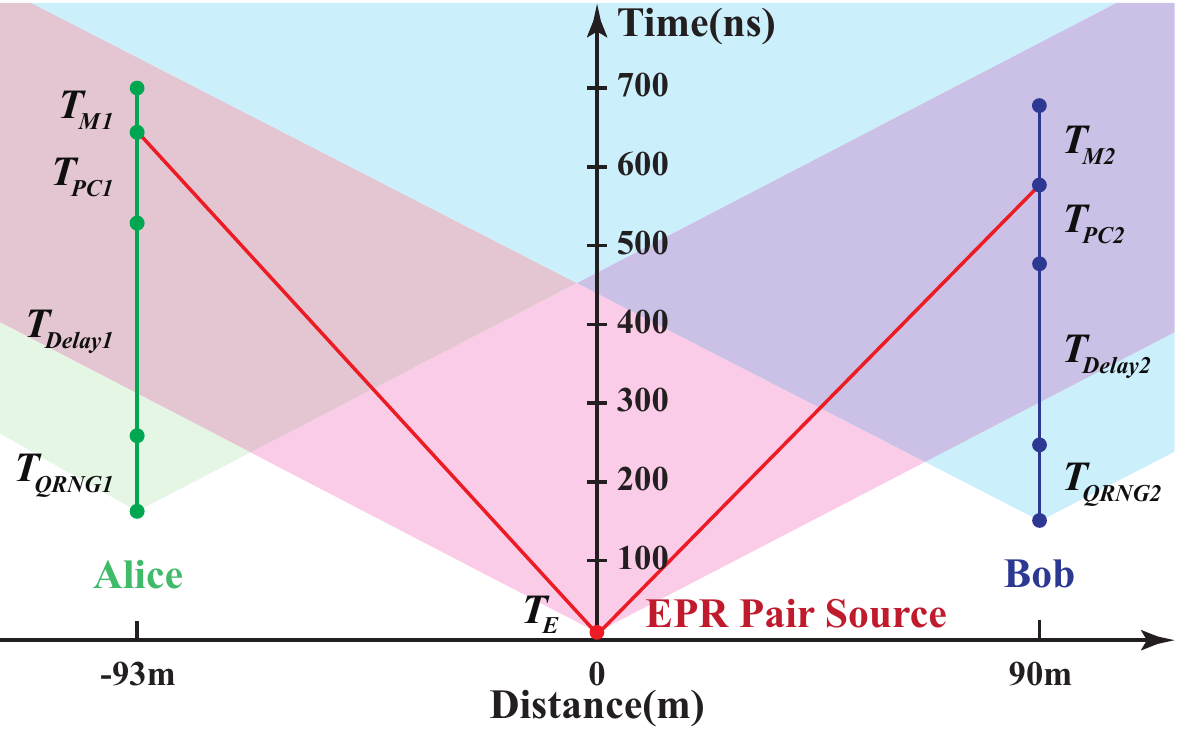}
    \caption{Spacetime configuration of the experiment}
    \label{fig:spacetime_detail}
\end{figure}
To close the locality loophole, space-like separation must be satisfied between relevant events at Alice and Bob’s measurement stations: the state measurement events by Alice and Bob, measurement event at one station and the setting choice event at the other station (Fig. 2(b) in the Main Text). We then obtain 
\begin{equation}
\left\{\begin{array}{l}
(|S A|+|S B|) / c>T_{E}-\left(L_{S A}-L_{S B}\right) / c+T_{Q R N G 1}+T_{\text {Delay } 1}+T_{P C 1}+T_{M 2} \\
(|S A|+|S B|) / c>T_{E}+\left(L_{S A}-L_{S B}\right) / c+T_{Q R N G 2}+T_{\text {Delay } 2}+T_{P C 2}+T_{M 1}
\end{array}\right.
\end{equation}
where $|SA|$ = 93 m ($|SB|$ = 90 m) is the free space distance between entanglement source and Alice’s (Bob’s)
measurement station, $T_E$ = 10 ns is the generation time for entangled photon pairs, which is mainly contributed by the 10 ns pump pulse duration, $L_{SA}$ = 191 m ($L_{SB}$ = 170 m) is the effective optical path which is mainly contributed by the long fibre (130 m, 116 m) between source and Alice/Bob’s measurement
station, $T_{QRNG1}$ = $T_{QRNG2}$ = 96 ns is the time elapse for QRNG to generate a random bit, $T_{Delay1}$ = 270 ns ($T_{Delay2}$ = 230 ns) is the delay between QRNG and Pockels cells, $T_{PC1}$ = 112 ns ($T_{PC2}$ = 100 ns) including the internal delay of the Pockcels Cells (62 ns, 50 ns) and the time for Pockcels cell to stabilize before performing single photon polarization state projection after switching which is 50 ns, which implies that the experimental time is able to be shortened by increasing the repetition rate of the experiment because the small q reduces the impact of the modulation rate of the Pockels cells, $T_{M1}$ = 55 ns ($T_{M2}$ = 100 ns) is the time elapse for SNSPD to output an electronic signal, including the delay due to fibre and cable length. $T_{R1,2}$ in Fig. 2(b) in the Main Text is the sum of $T_{QRNG1,2}$, $T_{Delay1,2}$ and $T_{PC1,2}$.

Measurement independence requirement is satisfied by space-like separation between entangled-pair creation event
and setting choice events, so we can have
\begin{equation}
\left\{\begin{array}{l}
|S A| / c>L_{S A} / c-T_{\text {Delay } 1}-T_{P C 1} \\
|S B| / c>L_{S B} / c-T_{\text {Delay } 2}-T_{P C 2}
\end{array}\right.
\end{equation}

\subsection{Experimental Results}
\subsubsection{Randomness Generation for Beacon}
To meet the require of randomness beacon, we shall generate at least 512 near-uniform random bits in one minute which is one period of randomness beacon. We set the soundness error $\varepsilon_h=2^{-64}$, $\varepsilon_x=2^{-100}$, the success probability for an honest protocol $\gamma=99.3\%$, and the lower bound to the success probability in the actual experiment to be the same as the soundness error in randomness expansion, $\kappa = \varepsilon_h=2^{-64}$. 

We obtain a set empirical input-output counts from a series of training trials, as shown in Table~\ref{tab:training_count}. The empirical frequency has a weak signalling behaviour due to the finite size effects. After applying a maximum likelihood estimation to the raw data, we obtain an empirical probability distribution $\nu(ABXY)$ from the training trials as shown in Table \ref{tab:training_probability}, which represents an expected experimental behaviour in randomness generation. With respect to this probability distribution, we follow the same approach in~\cite{li2021experimental} to derive a set of valid QEFs. Here, we summarise the final results. We first optimise a set of probability estimation factors (PEF) under the no-signaling condition and Tsirelson’s bounds. The power $\alpha$ of the optimized PEF is $\alpha = 1.0071$, and the values of the PEF are given in Table \ref{tab:PEF}. Such an optimisation problem is tackled via the parallel computation toolbox in Matlab. Afterwards, we rescale the PEF with a factor to obtain a valid QEF. The overall QEF rescaling factor is the multiplication of the sum of the 16 PEF values and $f_{max}$. We derived a valid value of $1+2.99\times 10^{-7}$ to the overall rescaling factor. The expected output entropy rate witnessed by the QEF is $r_\nu(F,\alpha) = 0.00398$. With these parameters, we determine the largest allowed number of trials in randomness expansion to be $N = 9.64 \times 10^6$ which takes about 48.2 seconds with the repetition frequency of 200kHz, and the success threshold in the experiment to be $h_s = 2.84\times 10^4$ bits.

\begin{table}
    \centering
    \caption{The empirical input-output counts from the training set.}
    \begin{tabular}{c|c c c c}
    \hline
    \diagbox[]{$(x,y)$}{$(a,b)$}&00 &01 &10 &11\\
    \hline
    00 & 1432368514	& 14021025	& 12251485	& 38365952\\
    01 & 1395132966	& 51261179	& 13585081	& 37027619\\
    10 & 1392797496	& 13586501	& 51825439	& 38798042\\
    11 & 1326545167	& 79804074	& 82131840	& 8497620\\
    \hline
    \end{tabular}
    \label{tab:training_count}
\end{table}

\begin{table}
    \centering
    \caption{The empirical input-conditioned probability distribution $\nu(AB|XY)$. A maximum likelihood estimation is applied to the empirical data to derive a probability distribution adapted to the model used. Here we present the result to 20 decimal places.}
    \begin{tabular}{c|c c c c}
    \hline
    \diagbox[]{$(x,y)$}{$(a,b)$}&00 &01 &10 &11\\
    \hline
    00 & 0.95682221443247694737	& 0.00936602447175378418	& 0.00818383944336322118	& 0.02562792165240606462\\	
    01 & 0.93194313970148556781	& 0.03424509920274514813	& 0.00907494837494691835	& 0.02473681272082236746\\	
    10 & 0.93038528791661345707	& 0.00907576916500537831	& 0.03462076595922665423	& 0.02591817695915446876\\
    11 & 0.88615498464810593671	& 0.05330607243351297847	& 0.05486310342832659281	& 0.00567583949005453364\\
    \hline
    \end{tabular}
    \label{tab:training_probability}
\end{table}

\begin{table}
    \centering
    \caption{The optimal PEF $F'(ABXY)$ with power $\alpha = 1.0071$. Here we present the result to 20 decimal places.}
    \begin{tabular}{c|c c c c}
    \hline
    \diagbox[]{(x,y)}{(a,b)}&00 &01 &10 &11\\
    \hline
    00 & 1.00022261334798057142 &   0.96652962358942828835 &	0.96475845132959570094 &	1.01624555865876731175\\ 
    01 & 1.00030376388124353504 &	0.98863956250689444261 &	0.95058782881190451164 &	1.02410372452527931308\\ 
    10 & 1.00032392821147841921 &	0.94883682088230669738 &	0.98897461161237376626 &	1.02266760137092616034\\ 
    11 & 0.99914969455929725228 &	1.01081389593364656676 &	1.01049901115840201626 &	0.93698311543584200667\\
    \hline
    \end{tabular}
    \label{tab:PEF}
\end{table}

\subsubsection{System Robustness}

\begin{figure}
    \centering
    \includegraphics[width = 0.9\textwidth]{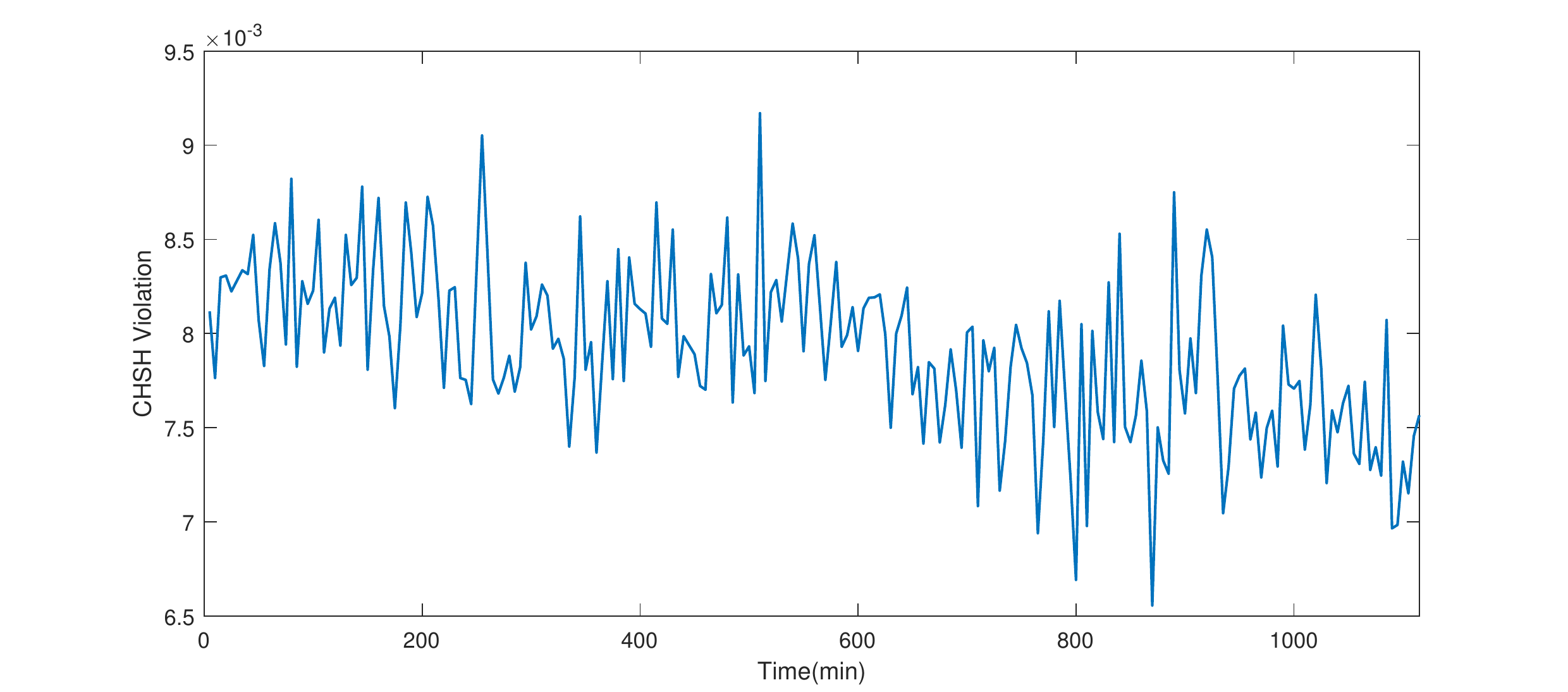}
    \caption{CHSH violation versus Time. Every point represents the average of the real-time CHSH violation of 5 minutes.}
    \label{fig:CHSH violation}
\end{figure}

We monitor the CHSH value during the generating process to represent the state of our experimental setup. We calculate the average CHSH violation every 5 minutes, as shown in Fig. \ref{fig:CHSH violation}. The mean value of the CHSH violation during the whole experiment of 1115 periods is 2.0079.

\subsubsection{Results for Repeated Randomness Generation Experiments}

\begin{figure}
    \centering
    \includegraphics[width = 0.9\textwidth]{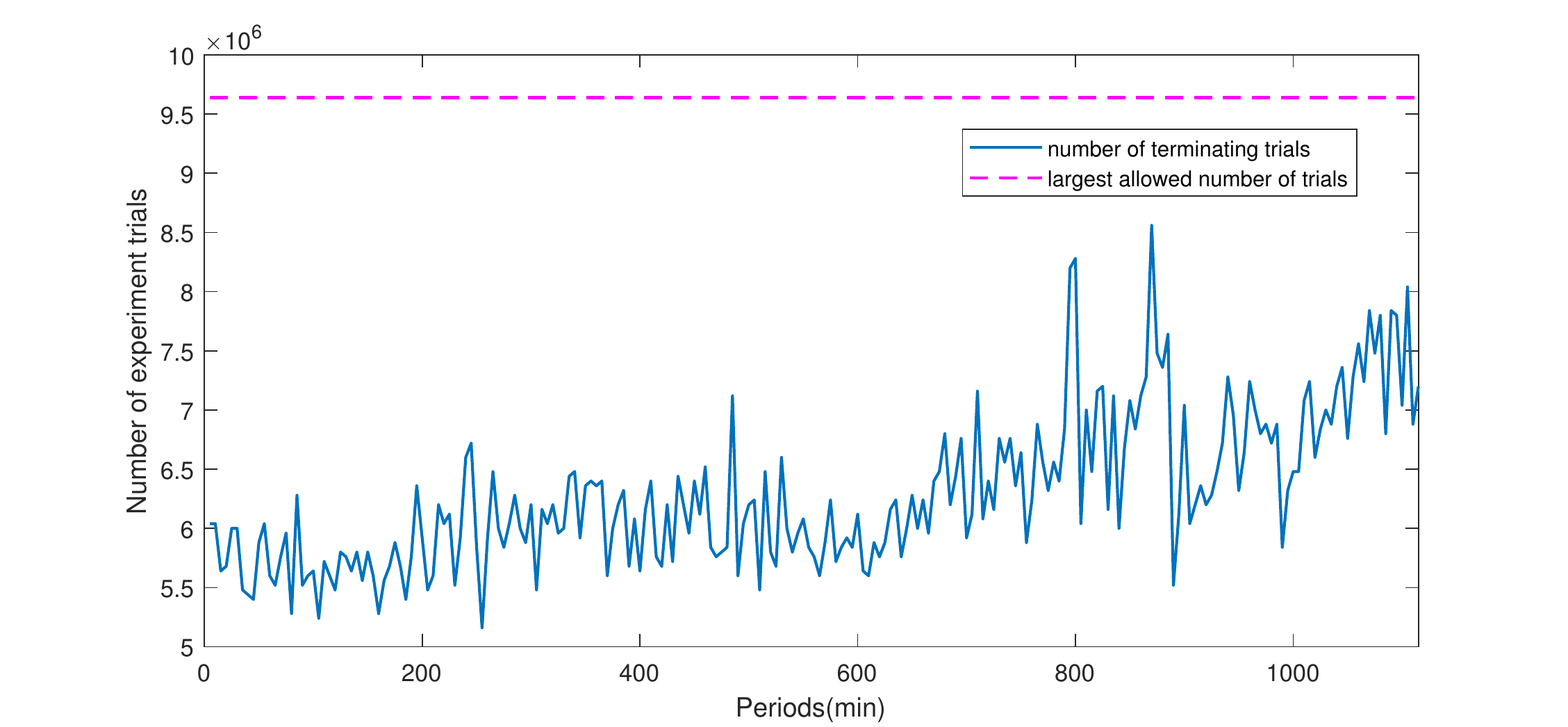}
    \caption{\textbf{Numbers of terminating trials in 1115 periods.} In each period, we loaded approximate $2 \times 10^5$ trials of data from time-digital converter (TDC) every second and updated the register $G_n$. The numbers of terminating trials is the number of trials when the register $G_n$ surpasses the success threshold. The largest allowed number of trials is shown as the dashed line. Because all of the number of terminating trials is below the largest allowed number, it shows the time is short enough that requires to generate 512bits randomness with error $\varepsilon_s=2^{-64}$ according to QPE protocol.}
    \label{fig:terminating trials}
\end{figure}

Our system is noise-resistant, which is capable of serving the randomness beacon for a long time stably. We repeat the randomness generating experiment for 1115 beacon periods and successfully obtain 512 bits of fresh randomness in every period. The numbers of experiment terminating trials are shown in Figure.\ref{fig:terminating trials}. The average time to terminate the experiment is 31.43 seconds, with a variance of 4.43 seconds, indicating that our DIQRNG source has low latency and is sufficiently consistent. We employ the hashing matrix to extract randomness from the raw data with failure probability parameters $\varepsilon_x=2^{100}$ as soon as the randomness generation process is stopped\cite{ma2013postprocessing}. Thanks to the use of fast Fourier transformation, the extraction procedure can be completed in approximately 2 seconds. Therefore, the entire time span for generating enough uniformly distributed quantum random numbers is short enough to match the practical requirements of the randomness beacon service.

\normalem

\end{document}